\documentclass[journal,twoside]{ieeecolor} 
\usepackage{tmi}
\usepackage{cite}
\usepackage{amsmath,amssymb,amsfonts}
\usepackage{algorithmic}
\usepackage{graphicx}
\usepackage{textcomp}
\def\BibTeX{{\rm B\kern-.05em{\sc i\kern-.025em b}\kern-.08em
    T\kern-.1667em\lower.7ex\hbox{E}\kern-.125emX}}
\usepackage{subfigure}
\usepackage{bm}
\usepackage{adjustbox}
\usepackage[justification=centering]{caption}
\usepackage{hyperref}
\hypersetup{
    colorlinks=true,
    linkcolor=blue,
    filecolor=magenta,      
    urlcolor=cyan,
    }

\newcommand\T{\rule{0pt}{2.6ex}}       
\newcommand\B{\rule[-1.2ex]{0pt}{0pt}} 

\def\r{\mathbf{r}}
\def\f{\mathbf{f}}
\def\H{\mathcal{H}}
\def\g{\mathbf{g}}
\def\L2{\mathbb{L}_2}
\def\R{\mathbb{R}}
\def\E{\mathbb{E}}
\def\n{\mathbf{n}}
\def\th{\mathrm{th}}

\begin{document}
\onecolumn
This manuscript has been accepted for publication on IEEE Transactions on Radiation and Plasma Medical Sciences (IEEE TRPMS) on May 03, 2022. Please use the corresponding reference when citing the manuscript.
\twocolumn
\newpage
\bstctlcite{IEEEexample:BSTcontrol}
\title{A projection-domain low-count quantitative SPECT method for $\alpha$-particle emitting radiopharmaceutical therapy}

\author{Zekun Li, Nadia Benabdallah, Diane S. Abou, Brian C. Baumann, Farrokh Dehdashti, David H. Ballard, Jonathan Liu,
Uday Jammalamadaka, Richard Laforest, Richard L. Wahl, Daniel L. J. Thorek, Abhinav K. Jha$^*$
\thanks{This work did not involve human subjects or animals in its research.}
\thanks{Z. Li is with the
Department of Biomedical Engineering, Washington University, St. Louis, MO 63130 USA.}
\thanks{N. Benabdallah, D. S. Abou, F. Dehdashti, D. H. Ballard, J. Liu, U. Jammalamadaka, R. Laforest, and R. L. Wahl are with the Mallinckrodt Institute of Radiology, Washington University, St. Louis, MO 63110, USA}
\thanks{B. C. Baumann is with the Department of Radiation Oncology, Washington University, St. Louis, MO 63110 USA.}
\thanks{D. L. J. Thorek is with the Department of Biomedical Engineering, Mallinckrodt Institute of Radiology, and Program in Quantitative Molecular Therapeutics, Washington University, St. Louis, MO, 63130 USA }
\thanks{*A. K. Jha is with the Department of Biomedical Engineering and Mallinckrodt Institute of Radiology, Washington University, St. Louis, MO 63130 USA (e-mail: a.jha@wustl.edu).}
}

\maketitle
\begin{abstract}
Single-photon emission computed tomography (SPECT) provides a mechanism to estimate regional isotope uptake in lesions and at-risk organs after administration of $\alpha$-particle-emitting radiopharmaceutical therapies ($\alpha$-RPTs). However, this estimation task is challenging due to the complex emission spectra, the very low number of detected counts ($\sim$20 times lower than in conventional SPECT), the impact of stray-radiation-related noise at these low counts, and the multiple image-degrading processes in SPECT. The conventional reconstruction-based quantification methods are observed to be erroneous for $\alpha$-RPT SPECT. To address these challenges, we developed a low-count quantitative SPECT (LC-QSPECT) method that directly estimates the regional activity uptake from the projection data (obviating the reconstruction step), compensates for stray-radiation-related noise, and accounts for the radioisotope and SPECT physics, including the isotope spectra, scatter, attenuation, and collimator-detector response, using a Monte Carlo-based approach. The method was validated in the context of three-dimensional SPECT with $\bm{\mathrm{^{223}Ra}}$, a commonly used radionuclide for $\alpha$-RPT. Validation was performed using both realistic simulation studies, including a virtual clinical trial, and synthetic and 3D-printed anthropomorphic physical-phantom studies. Across all studies, the LC-QSPECT method yielded reliable regional-uptake estimates and outperformed the conventional ordered subset expectation maximization (OSEM)-based reconstruction and geometric transfer matrix (GTM)-based post-reconstruction partial-volume compensation methods. Further, the method yielded reliable uptake across different lesion sizes, contrasts, and different levels of intra-lesion heterogeneity. Additionally, the variance of the estimated uptake approached the Cram\'er-Rao bound-defined theoretical limit. In conclusion, the proposed LC-QSPECT method demonstrated the ability to perform reliable quantification for $\alpha$-RPT SPECT.
\end{abstract}


\section{Introduction}
\label{sec:introduction}
Targeted radionuclide therapy with $\alpha$-particle-emitting therapeutic isotopes is gaining increasing clinical significance. 
Many potent $\alpha$-particle radiopharmaceutical therapies ($\alpha$-RPTs), including those based on Radium-223~\cite{kluetz2014radium}, Actinium-225~\cite{kratochwil2017targeted,mcdevitt2018feed}, Bismuth-213~\cite{jurcic2002targeted}, and Astatine-211~\cite{zalutsky2008clinical}, are under pre- and clinical investigation for their ability to ablate tumors while minimizing damage to the surrounding normal tissues~\cite{RN39, RN40}.
These isotopes distribute throughout the patient, accumulating to unknown levels at sites of disease and in radiosensitive vital organs. Thus, methods to quantify absorbed doses in lesions and at-risk organs are much needed to adapt treatment regimens, predict therapy outcomes, and monitor adverse events~\cite{brans2007clinical}.
Multiple studies~\cite{garin2020personalised, siegel2003red, RN69} show the benefits of such dose quantification in personalizing RPT regimens.

Often, the $\alpha$-particle-emitting isotopes also emit $\gamma$-ray photons that can be detected by a $\gamma$ camera. This provides a mechanism to quantify the absorbed dose from the regional activity uptake in organs and lesions.
Currently, this regional uptake quantification has been explored using planar~\cite{RN65} and single-photon emission computed tomography (SPECT)-based methods~\cite{RN21,RN60,RN61}. Of these, the planar-based methods are known to suffer from inaccuracy caused by organ overlap and background activity~\cite{RN18}. SPECT is tomographic and is thus less affected by this source of inaccuracy.

Contemporary methods to quantify regional uptake by SPECT first require reconstruction of the activity distribution over a voxelized grid. Next, a volume of interest (VOI) is defined over this grid, corresponding to the region over which the uptake is desired. The regional uptake is estimated by averaging the activity over all the voxels within this VOI~\cite{RN21,RN60,RN61}.
However, these methods are challenged at low number of detected counts, as is the case in $\alpha$-RPTs, where the administered activity is up to 1000 times lower than conventional radionuclide therapies. For example, when imaging with $\mathrm{^{223}Ra}$, the number of detected counts can be as low as 5,000 counts per axial slice in the projection domain. This is multiple folds lower than conventional SPECT studies and results in high bias in estimated activity (19\%-35\%) even with highly fine-tuned protocols~\cite{RN21,RN60,RN61}. Another major concern is the limited precision in the estimated activity with these methods at low count levels~\cite{RN61}. 
Thus, there is an important need for improved methods to accurately and precisely quantify regional uptake from SPECT measurements for $\alpha$-RPTs. 
	
We note here that reconstruction is only an intermediary step for quantification. Reconstruction requires estimating activity in a large number of voxels, an ill-posed problem that becomes even more challenging when the number of detected counts is small~\cite{RN50}.
In this context, we recognize that the number of VOIs over which the regional uptake needs to be estimated is far fewer than the number of voxels. 
Thus, directly quantifying mean uptake in these VOIs from the projection data is a less ill-posed problem.
Methods for this purpose have been reported~\cite{RN19} 
and have demonstrated improved performance compared to reconstruction-based quantification methods~\cite{RN19,jha2015estimating,RN71,RN72,RN62,moore2011improved,southekal2012evaluation}. 
However, such projection-domain quantification methods have been developed for conventional SPECT studies. At low counts, in addition to the poor signal-to-noise ratio, effects such as stray-radiation-related noise due to photons emitted from regions other than the patient become significant. 
Our studies, as presented later, show that not accounting for this noise leads to highly erroneous activity estimates. 
Further, $\alpha$-particle-emitting isotopes typically have complicated emission spectra. Additionally, image-degrading processes such as attenuation, scatter, septal penetration, and finite energy and position resolution of the detector make the quantification task even more challenging.

To address the above-mentioned challenges, we propose a low-count quantitative SPECT (LC-QSPECT) method that advances the idea of direct quantification from projection data while incorporating the modeling of stray-radiation-related noise and the imaging physics and spectra of $\alpha$-RPT isotopes. Our initial studies based on this idea were presented as an abstract~\cite{li2021projection}. In this manuscript, we provide a significantly more detailed treatment of the method in terms of theory, evaluation, and analysis. We first provide the theoretical foundations of the method.

\section{Theory}
Consider a SPECT system imaging a radioisotope distribution $f(\r)$, where $\r = (x,y,z)$ denote the spatial 3D coordinates.
Denote the measured projection data by the $M$-dimensional vector $\g$.
Assume that the object being imaged and the projection data lie in the Hilbert space of square-integrable functions, denoted by $\L2(\R^3)$ and the Hilbert space of Euclidean vectors, denoted by $\E^M$, respectively.
Then, the SPECT system, denoted by the operator $\H$, is a transformation from $\L2(R^3)$ to $\E^M$.
In SPECT with $\alpha$-RPTs, the stray-radiation-related noise occupies a substantial portion of the measured counts due to the very low-count levels. 
We model this noise as Poisson distributed with the same mean $\psi$ for all projection bins.
Let $\Psi$ be an $M$-dimensional vector with each element equal to $\psi$ that denotes the mean stray-radiation-related noise across all $M$ projection bins.
Denote the entire noise in the imaging system by the $M$-dimensional random vector $\n$. Then the projection data $\bm{g}$ are Poisson distributed with mean $\H \f + \Psi$.
Thus, the imaging system equation is given by 
\begin{equation}
\g = \H \f + \Psi+ \n .
\label{g}
\end{equation}
Our objective is to estimate the regional uptake within a set of VOIs. Mathematically, we first define a 3D VOI function $\phi_k^{VOI}(\bm{r})$, where
\begin{equation}
\label{eq:dif_roi}
\phi^{VOI}_{k}(\bm{r}) =\left\{
        \begin{array}{cc}
             1,&{\mathrm{if~}\bm{r}\mathrm{~lies~within~the~k^{th}~VOI.}} \\
             0,&  {\mathrm{otherwise}}.
        \end{array}
\right.
\end{equation}
Denote $\bm{\lambda}$ as the $K$-dimensional vector of regional uptake, where $\lambda_k$ is given by
\begin{equation}
\label{eq:roi_apply}
\lambda_k = \frac{\int d^3 r f(\r) \phi_k^{VOI}(\r)}{\int d^3 r \phi_k^{VOI}(\r)}.
\end{equation}
Our objective is to estimate $\bm{\lambda}$.

\subsection{Reconstruction-based quantification (RBQ) methods}
\label{sec:Rec_based_qtf}
The conventional procedure to estimate $\boldsymbol{\lambda}$ is to first reconstruct the activity uptake distribution over a voxelized grid, and then estimate the activity uptake in a discretized version of the VOI as defined in Eq.~\eqref{eq:dif_roi}.
Mathematically, the activity uptake distribution is described using a voxel basis function denoted by $\phi_n^{vox}(\r)$ as
\begin{equation}
f_{vox}(\r)  = \sum_{n=1}^N \theta_n \phi_n^{vox}(\r).
\end{equation}
Multiple algorithms are available to estimate the coefficients $\theta_n$.
These algorithms yield an estimate of $\theta_n$, denoted by $\hat{\theta}_n$. To estimate $\lambda_k$, we first define a discretized mask matrix $\mathbf{M}$.
One procedure to define the elements of this matrix is given by
\begin{equation}
\label{eq:VOI_def}
M_{n,k} = \left\{
        \begin{array}{cc}
        1, &{\mathrm{if~a~majority~of~the~n^{th}~voxel}}\\
           & {\mathrm{lies~inside~the~k^{th}~VOI.}} \\
        0, &{\mathrm{otherwise}}.
        \end{array}
\right.
\end{equation}
Then, the estimate of $\lambda_k$ obtained from the reconstructed image, denoted by $\hat{\lambda}_k^{recon}$, is given by
\begin{equation}
\hat{\lambda}_k^{recon} = \frac{1}{N} \sum_{n=1}^N \hat{\theta}_n M_{n,k}.
\label{eq:est_lambda_theta}
\end{equation}
This procedure to estimate $\bm{\lambda}$ has several issues.
First, a large number of voxels need to be estimated during reconstruction, leading to a highly ill-posed problem, especially when the number of counts is low, and leading to biased estimate~\cite{cloquet2012mlem}.
A second issue is the bias introduced due to partial volume effects (PVEs)~\cite{RN44}. PVEs include two distinct phenomena. The first is due to the finite system resolution. Another is the tissue-fraction effects. More specifically, when defining $M_{n,k}$, an element of this matrix is $1$ when a majority of this voxel is within the VOI. Therefore, this does not define a continuous VOI, causing bias when estimating $\hat{\lambda}_k^{recon}$ using Eq.~\eqref{eq:est_lambda_theta}.
The third issue is the activity inside a voxel $\theta_n$ is fundamentally not estimable~\cite{lehovich2005list}.
Next, by the data-processing inequality, the process of reconstruction can only lead to information loss~\cite{RN50,beaudry2011intuitive}.
Finally, these RBQ approaches are often based on maximum-likelihood expectation maximization (MLEM)~\cite{shepp1982maximum} or ordered subset expectation maximization (OSEM)~\cite{hudson1994accelerated}. However, at low counts, these methods have limited precision and deviate from the theoretically lowest possible Cram\'er-Rao bound (CRB)~\cite{cloquet2012mlem}.
All these issues serve as sources of error in the estimated regional activity.
Thus, as previous studies have reported, even highly fine-tuned versions of these methods yield unreliable estimates of regional uptake~\cite{RN21,RN60,RN61}.

\subsection{Proposed method}
To address the above-described issues with RBQ methods, we recognize that our objective is to estimate the mean uptake within certain regions, $\bm{\lambda}$. 
Thus, we directly represent the object $f(\r)$ in terms of the VOI-basis functions. These VOI basis functions are given by $\phi^{VOI}_{k}(\bm{r})$ as defined in Eq.~\eqref{eq:dif_roi}. The activity distribution is then represented in terms of these basis functions as
\begin{equation}
\label{eq:roi_rep}
f_{VOI} (\r ) = \sum_{k=1}^K \lambda_k \phi_k^{VOI}(\r),
\end{equation}
where, if the activity inside each VOI is constant, then $f_{VOI}(\r) = f(\r)$.
Inserting this definition for $f(\r)$ in Eq.~(\ref{g}) yields the following expression for the $m^{\th}$ element of the vector $\g$
\begin{equation}
    \begin{split}
    g_m &= \int h_m(r)f(r)d^{3}r + \psi + n_m \\
        &= \sum_{k=1}^{K} \lambda_{k}\int h_m(r)\phi_{k}^{VOI}(r)d^{3}r+ \psi +n_m.
    \end{split}
\end{equation}
This can be written in vector form as
\begin{equation}
\label{eq:gnoise}
    \g = \bm{H} \bm{\lambda} + \Psi +\n,
\end{equation}
where $\bm{H}$ is the $M~\times~K$ dimensional system matrix with elements given by
\begin{equation}
H_{mk} = \int d^3 r h_m(\r) \phi_k^{VOI} (\r).
\end{equation}
Given the measured projection $\g$, to estimate $\bm{\lambda}$, we maximize the probability of occurrence of the measured data. Denote $\Pr(x)$ as the probability of a discrete random variable $x$. Then, the probability of the measured projection data is given by
\begin{equation}
    \Pr(\g|\bm{\lambda}) = \prod_{m=1}^M \Pr(g_m |\bm{\lambda}), 
\end{equation}
where we have used the fact that the measured data across the different bins are independent. Now, the measured data $g_m$ is Poisson distributed with mean $(\bm{H \lambda})_m + \psi$. Thus:
\begin{equation}
    \Pr(\g|\bm{\lambda}) = \prod_{m=1}^{M}\exp[-(\bm{H}\bm{\lambda})_m-\psi]\frac{[(\bm{H}\bm{\lambda})_m+\psi]^{g_m}}{{g_m}!}.
    \label{eq:poisson}
\end{equation}
This gives the likelihood of the measured data $\g$.
To estimate $\bm{\lambda}$, we maximize the logarithm of the likelihood of $\bm{\lambda}$ given $\g$
\begin{equation}
    \widehat{\bm{\lambda }}=\arg \underset{\bm{\lambda }}{\mathop{\max }}\,\ln [\Pr (\mathbf{g}|\bm{\lambda })].
    \label{optGoal}
\end{equation}
To maximize this log-likelihood, we follow the same process as used to derive the conventional MLEM technique~\cite{RN50}. Briefly, we differentiate the log-likelihood with respect to the elements of $\bm{\lambda}$ and equate that to $0$ to find the point at which the log-likelihood is maximized.
This yields the following iterative equation to estimate $\lambda_k$:
\begin{equation}
   \hat{\lambda}_{k}^{(t+1)} = \hat{\lambda}_{k}^{(t)}\frac{1}{\sum\limits_{m=1}^{M} H_{mk}}
   \sum\limits_{m=1}^{M}
   \frac{g_{m}}{[\bm{H\hat{\lambda}}^{(t)}]_{m}+\psi} H_{mk},
   \label{eq:MLEM}
\end{equation}
where $\hat{\lambda}_{k}^{(t)}$ denotes the estimate of $\lambda_k$ at the $t^{th}$ iteration. We refer to this procedure as the low-count quantitative SPECT (LC-QSPECT) method. 
Our approach advances on existing methods that directly quantify from projection data~\cite{RN19} by providing the ability to model stray-radiation-related noise. As we see later, this ability to model stray-radiation-related noise plays a key role in the task of reliable quantification. 
Further, the system matrix $\bm{H}$ models all key image-degrading processes in $\alpha$-particle SPECT using a Monte Carlo (MC)-based approach, which further improves the performance of this technique on the task of reliable quantification. 

Our approach alleviates the issues outlined earlier with RBQ approaches.
Typically, the number of VOIs $K$ is less than the number of voxels $N$, so the problem is less ill-posed. Besides, the method is less sensitive to PVEs since we define the boundaries of VOIs before estimating the regional uptake.
In particular, the tissue-fraction effects are minimized since there is no voxelization. Further, while it is true that the mean VOI activity $\lambda_k$ could also be inestimable, but since the VOI is generally larger than a voxel, the estimation bias is lower. We also directly estimate the regional uptake from the projection data, thus avoiding any reconstruction-related information loss. Finally, as our results later show, the method yields estimates with a precision that is close to the CRB.


\section{Implementation and evaluation of the proposed method}
\subsection{Implementation}
\label{sec:implementation}
Implementing the proposed LC-QSPECT method required obtaining the elements of the system matrix $H_{mk}$, as shown in Eq.~\eqref{eq:MLEM}. 
We obtained these elements using an MC-based approach. More specifically, we used SIMIND, a well-validated MC-based simulation software~\cite{ljungberg1989monte,RN46} to model the isotope emission and the SPECT system. 
Next, for a given patient, we obtained the definition of the VOIs.
These can be obtained, for example, by segmenting the CT that is acquired along with the SPECT. 
We assigned unit uptake to the VOI and zero uptake elsewhere. We assumed that the attenuation map of the patient is available. Projection data for this activity and attenuation map were generated by simulating more than 100 million photons for each VOI. 
The simulations modeled all relevant image-degrading processes in SPECT including attenuation, scatter, collimator response, septal penetration and scatter, characteristic X-ray from both the $\alpha$-emitting isotope and the lead in the collimator, finite energy and position resolution of the detector, and the backscatter in the detector.
Scaling the projection data according to the acquisition time of the projections yielded the corresponding columns of the system matrix.
While the bremsstrahlung was not modeled in this study, this does not significantly affect the accuracy of the simulation, as we show in Sec.~\ref{sec:res_realism}.

Next, the LC-QSPECT method required obtaining the mean of stray-radiation-related noise, i.e. $\psi$ in Eq.~\eqref{eq:MLEM}. We estimated this experimentally from a planar blank scan acquired on the SPECT system for over 10~minutes. Averaging the projection bin counts in this scan yielded the mean background counts. This was then scaled to the acquisition time to estimate the mean stray-radiation-related noise. 

The computed system matrix and mean stray-radiation-related noise were used in Eq.~\eqref{eq:MLEM} to estimate the regional uptake directly from the projection data. 
As the system matrix modeled all relevant image-degrading processes, these processes were automatically compensated during quantification.

\subsection{General evaluation framework}
Evaluating the performance of the LC-QSPECT method on the estimation task of regional uptake quantification required a setup where the ground-truth regional uptake was known. For this purpose, we conducted realistic simulations, including a virtual clinical trial, and physical-phantom studies. We describe these evaluations in detail in Sec.~\ref{sec:design_simu}. In this sub-section, we describe the methods to which we compared our method and the figures of merit.

\subsubsection{Methods compared}
\label{sec:methods_compared}
We compared the performance of the LC-QSPECT method with two widely used RBQ methods.

a. OSEM-based method: Here, the activity maps were first reconstructed using an OSEM-based approach.
This approach, implemented using the Customizable and Advanced Software for Tomographic Reconstruction (CASToR)~\cite{RN49} software, compensated for attenuation, scatter, collimator-detector response, and stray-radiation-related noise. 
Scatter was compensated using the triple-energy-window (TEW) method~\cite{RN47}.
To minimize noise amplification that could be caused by using the TEW method, we applied pre-reconstruction Butterworth filters to the photopeak and scatter-window projections~\cite{king1997investigation}. 
These filters were optimized by minimizing the normalized root mean square error (NRMSE) between the true and estimated regional uptake values.
The cut-off frequencies of the filters applied to photopeak and scatter-window projections were optimized by a two-dimensional grid search and the optimized frequencies were found to be 0.15~cycle/pixel and 0.05~cycle/pixel, respectively. The optimized orders were found to be 8 for both filters. 
We also compensated the stray-radiation-related noise with an additive term in the iteration equation of the OSEM-based method similar to that of the proposed method as described in Sec.~\ref{sec:implementation}.
The reconstructed-image dimensions were 128 $\times$ 128 $\times$ 91, with a voxel side-length of 4.418~mm. 
We also optimized the number of iterations and subsets based on the NRMSE between the true and estimated regional uptake.
The optimized number of iterations and subsets were found to be 20 and 6, respectively, which was consistent with that reported in~\cite{RN61}.
From the reconstructed image, the uptake in different VOIs were calculated.

b. GTM-based method: 
PVEs are known to degrade quantification accuracy in SPECT~\cite{RN44}. The LC-QSPECT method implicitly assumes constant uptake within each VOI. Under this assumption, PVEs can also be compensated post-reconstruction.
A widely used approach for this purpose is the geometric transfer matrix (GTM)-based method~\cite{rousset1998correction}. Thus, we also compared our approach to this method.
The elements of the GTM were obtained from the projections and reconstructions of the VOIs. These projections were obtained using the process described in Sec.~\ref{sec:implementation} and the reconstruction was done using the OSEM-based method as described above.
The rest of the implementation of this method was as described in~\cite{rousset1998correction}.

\subsubsection{Figures of merit}
\label{sec:figure_of_merit}
We evaluated the accuracy, precision, and overall error of the LC-QSPECT, OSEM, and GTM-based methods on the task of estimating regional uptake. In most of our experiments, we generate multiple instances of projection data for a single phantom, where each instance corresponds to a separate noise realization. 
Denote the total number of noise realizations by $R$.
Denote the true and estimated activity uptake in the $k^{\th}$ VOI for the $r^{\th}$ noise realization by $\lambda_{rk}$ and $\hat{\lambda}_{rk}$, respectively.
In these experiments, the accuracy of the estimated uptake was quantified using the normalized bias (NB), which, for the $k^{th}$ VOI, is given by
\begin{equation}
\label{eq:NB}
    \mathrm{NB}_{k}=\frac{1}{R}
    \sum\limits_{r=1}^{R} \frac{\hat{{\lambda }}_{rk}-{{\lambda }_{rk}}}{{{\lambda }_{rk}}}.
\end{equation}
The precision of the estimated uptake was quantified using the normalized standard deviation (NSD), which, for the $k^{th}$ VOI, is given by
\begin{equation}
\label{eq:NSD}
    \mathrm{NSD}_{k} = \sqrt{
    \frac{1}{R-1}
    \sum\limits_{r=1}^{R}{\left(\frac{\hat{{\lambda }}_{rk}}{\lambda_{rk}}-\frac{1}{R}\sum\limits_{r'=1}^{R}\frac{\hat{{\lambda }}_{r'k}}{\lambda_{r'k}}\right)^{2}}
    }.
\end{equation}
Finally, the overall error in estimating the uptake was quantified by the NRMSE. For the $k^{th}$ VOI, this is given by
\begin{equation}
    \mathrm{NRMSE}_{k} = \sqrt{NB_k^2+NSD_k^2
    }.
\end{equation}
To evaluate the performance of the methods over populations, we used the ensemble NB and ensemble NRMSE.
Denote the number of samples in the population by $S$. Denote the true and estimated activity uptake in the $k^{\th}$ VOI for the $s^{\th}$ sample by $\lambda_{sk}$ and $\hat{\lambda}_{sk}$, respectively. 
The ensemble NB for the $k^{th}$ VOI is given by
\begin{equation}
    \mathrm{Ensemble~NB}_{k}=\frac{1}{S}
    \sum\limits_{s=1}^{S} \frac{\hat{{\lambda }}_{sk}-{{\lambda }_{sk}}}{{{\lambda }_{sk}}}.
\end{equation}
The ensemble NRMSE for the $k^{th}$ VOI is given by
\begin{equation}
    \mathrm{Ensemble~NRMSE}_{k} = \sqrt{\frac{1}{S}
    \sum\limits_{s=1}^{S} \left(\frac{\hat{{\lambda }}_{sk}-{{\lambda }_{sk}}}{{{\lambda }_{sk}}}\right)^2}
\end{equation}

Next, in cases where we had just a single noise realization, we used normalized error to quantify performance. This is defined as the difference between the true and estimated uptake values, normalized by the true uptake value.

Finally, we also computed the CRB, which is the minimum variance that can be achieved by an unbiased estimator, as a benchmark to compare the precision of the activity estimated using the proposed method.
The CRB is given by the diagonal elements of the inverse of the Fisher information matrix for the estimated parameter~\cite{RN50}. Denote this matrix by $\bm{F}$.
Denote $\lambda_{k}$ and $\hat{\lambda}_{k}$ as the true and estimated activity uptake, respectively, in the $k^{th}$ VOI of a targeted digital phantom. Then, the elements of this matrix are given by
\begin{equation}
    F_{k_{1}k_{2}}=-E\left[\frac{\partial^{2} }
    {\partial \lambda_{k_{1}} \partial \lambda_{k_{2}} }
    \ln \Pr (\bm{g}|\bm{\lambda})\right].
    \label{CRB1}
\end{equation}
Substituting Eq.~(\ref{eq:poisson}) in Eq.~\eqref{CRB1} yields
\begin{equation}
    F_{k_{1}k_{2}}=
    \sum\limits_{m=1}^{M}{\frac
    {H_{mk_1} H_{mk_2}}
    {
    (\bm{H} \bm{\lambda})_{m} + \psi
    }
    }.
    \label{CRB}
\end{equation}

\subsection{Evaluation using realistic simulation studies}
\label{sec:design_simu}
We conducted realistic simulation studies in the context of imaging patients with bone metastases of prostate cancer treated with $\mathrm{^{223}Ra}$, a widely administered United States (US) Food and Drug Administration (FDA)-approved $\alpha$-RPT indicated for the treatment of patients with castration-resistant prostate cancer~\cite{RN54,RN55}.
A major site of these osseous metastases is the pelvis, which is adjacent to the anatomical location of the prostate. Thus, we focused on the pelvic region.

\begin{figure}
    \centering
    \subfigure[]{
    \label{act_map}
    \includegraphics[width=0.23000\textwidth]{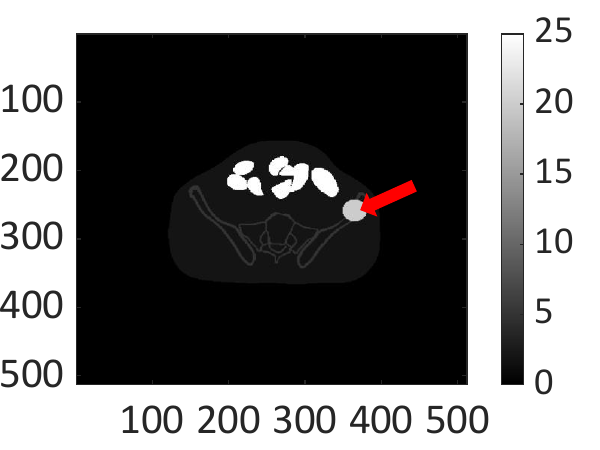}}
    \subfigure[]{
    \includegraphics[width=0.23000\textwidth]{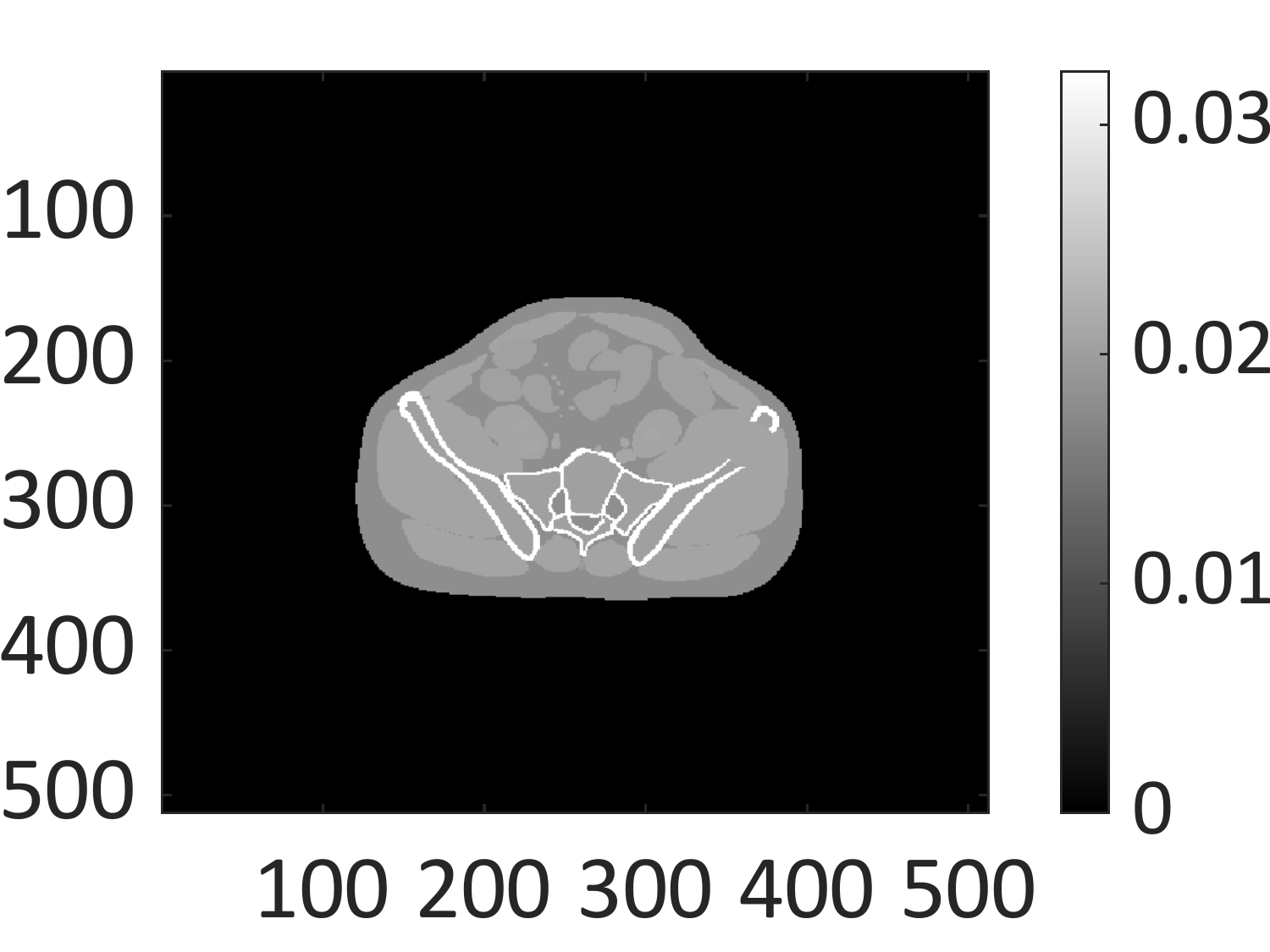}}
    \caption{The digital (a) activity map and (b) attenuation map for the pelvic region generated using the anthropomorphic XCAT phantom}
    \label{fig:phantomSlice}
\end{figure}

To generate a realistic patient population for our study, digital 3D activity and attenuation maps of the pelvic region were generated using the Extended Cardiac-Torso (XCAT)~\cite{RN45} phantom (representative slice shown in Fig.~\ref{fig:phantomSlice}). To simulate continuous activity distribution, the activity map had a high resolution of 512 $\times$ 512 along the axial dimensions and 364 slices along the depth dimension. The side length of the voxels was 1.105~mm. Our analysis of clinical SPECT data of patients administered $\mathrm{^{223}Ra}$ therapy suggested that there were three primary sites of uptake in the pelvic region: lesion (indicated by the arrow), bone, and gut (through which the majority of administered activity is cleared). The rest of the region in the patient typically had the same low uptake. We refer to the VOI corresponding to the rest of the patient as the background. This led to a total of four VOIs. The variability in anatomies and regional uptake within these phantoms were simulated based on clinical data, as described later (Sec.~\ref{sec:VCT}). 

Next, a GE Optima 640 SPECT system with a high energy general purpose (HEGP) collimator was simulated using SIMIND. 
The scintillation detector in the system had an intrinsic spatial resolution of 3.9~mm and an energy resolution of 9.8\% at 140~keV, where the resolutions were quantified in terms of full width at half maximum (FWHM). Photons were acquired at 60 angular positions spaced uniformly over 360$^{\circ}$.
The photopeak window was set as 85~keV~$\pm$~20\%~\cite{RN51}.
The isotope emission and all relevant image-degrading processes in SPECT were simulated. 
The mean of the stray-radiation-related noise in each projection bin, $\psi$, was determined as described in Sec.~\ref{sec:implementation}. The value of stray-radiation-related counts in each projection bin was individually sampled from a Poisson distribution with a mean equal to $\psi$ and was added to the corresponding projection bin in the simulated projections.
The generated projections had around 5,000 counts per axial slice to simulate a clinically realistic low-count $\alpha$-RPTs SPECT acquisition.
We used the procedure described in Sec.~\ref{sec:implementation} to quantify the uptake from the simulated projection data using the proposed, OSEM, and GTM-based methods. Then, the methods were evaluated on the task of quantifying the regional uptake using the figures of merit defined in Sec.~\ref{sec:figure_of_merit}.

\subsubsection{Validating the SPECT simulation}
We conducted our simulations with SIMIND. While SIMIND has been validated for multiple SPECT studies~\cite{RN46,morphis2021validation}, we further validate the accuracy of our simulation approach in the context of $\alpha$-RPT SPECT. For this purpose, we compared the projection data obtained with our simulation approach to that obtained on an actual scanner. 
More specifically, we considered a NEMA phantom that was scanned on a GE Optima 640 SPECT system with HEGP collimator using the procedure as described in more detail in Sec.~\ref{sec:phy_pht}. We then modeled this acquisition using our simulation approach. For our simulation, the activity map was designed to simulate the known $\mathrm{^{223}Ra}$ activity concentrations filled in the physical phantom and the attenuation map of the NEMA phantom was derived from the CT scans. Then we generated the simulated projections using SIMIND, modeling the same acquisition process as described in Sec.~\ref{sec:phantom_scan}. 
The profiles of the projection data at four angular positions spaced uniformly over 360$^{\circ}$ obtained with the scanner and that with the simulation approach were compared directly, without any normalization. 
The match of these profiles, as we see later in the results section, provided evidence of the accuracy of our simulations.

\subsubsection{Evaluating convergence of the LC-QSPECT method}
\label{sec:converge_speed}
For this purpose, we generated five 3D XCAT-based phantoms with dimensions similar to average patient size. Each phantom had a lesion of a different size but at the same location in the pelvis. The lesion diameters varied between 15~mm to 35~mm. 
The activity uptake in the four VOIs was in the ratio of 2:5:25:20 in
the background, bone, gut, and lesion regions. These ratios were derived based on our analysis of clinical data. 
Projection data corresponding to these phantoms were generated as described in Sec.~\ref{sec:design_simu}. We applied the LC-QSPECT method to the generated projections and the error in the estimated lesion uptake after each iteration was computed. 600 iterations were performed for each phantom.

\subsubsection{Evaluating performance in a simulated clinical scenario using a virtual clinical trial (VCT)}
\label{sec:VCT}
VCTs are an emerging evaluation paradigm that provides the ability to rigorously and objectively evaluate the performance of new imaging technology in simulated clinical scenarios that model patient population variability~\cite{abadi2020virtual}. 
In our VCT, we simulated 50 digital 3D male patients with different anatomies using the XCAT phantom. As in Segars~\cite{RN45}, the heights and weights of the 50 patients were sampled from a Gaussian distribution with the mean equal to the height and weight of a $50^{th}$ percentile male US adult and a 10\% standard deviation.
Next, based on clinically derived parameters~\cite{RN65},
the lesion diameter was sampled from a Gaussian distribution with a mean of 33.75~mm and a standard deviation of 12.64~mm.
Next, as described above, the activity uptake in the four VOIs were sampled from a normal distribution with a clinically derived mean uptake ratio of 2:5:25:20 in the background, bone, gut, and lesion regions.
Projection data corresponding to these patients were generated as described in Sec.~\ref{sec:design_simu}. The LC-QSPECT, OSEM, and GTM-based methods were applied to these data, thus comparing the performance of these methods for this realistic population.
We evaluated the accuracy and overall error of the regional uptake estimates yielded by the LC-QSPECT method across all these regions for this clinically realistic patient population.

\subsubsection{Comparing precision of estimated regional uptake using LC-QSPECT method with Cram\'er-Rao Bound}
\label{sec:CRB}
From our results, we observed that the LC-QSPECT method was yielding approximately unbiased estimates of the regional activity uptake. For an unbiased estimator, the minimum variance that can be obtained is given by the Cram\'er-Rao bound. Thus, we compared the precision of the estimated regional uptake with this bound. 
To conduct this study, we generated 50 noise realizations for a 3D male patient simulated using XCAT. The LC-QSPECT method was used to estimate the mean regional activity in each of the 50 noise realizations. The NB and NSD of these estimates were computed.
The NSD of estimates in each VOI was compared to the square root of the CRB.

\subsubsection{Evaluating performance as a function of lesion size and contrast}
\label{sec:effSize}

To evaluate the sensitivity of the method to lesion size, we used the same simulation setup as described in Sec.~\ref{sec:converge_speed}. 50 noise realizations were generated for each of the five phantoms, where each phantom had a different lesion size.
To evaluate the sensitivity of the method to lesion contrast, we recognize that the lesion is present within the bone. We generated six phantoms with average patient size and a lesion of diameter 33.75~mm~\cite{RN65} in the pelvis. 
Each phantom had a different lesion-to-bone uptake ratio (LBUR), ranging from 1:1 to 6:1.
The uptake in the background, bone, and gut regions was in the ratio of 2:5:25.
50 noise realizations were generated for each phantom and used to
evaluate the method performance as a function of lesion contrast.

\subsubsection{Evaluating effect of spatial intra-lesion heterogeneity}
In contrast to the OSEM-based method, the LC-QSPECT method assumes that the activity uptake in the VOI is homogeneous. This assumption may not hold in clinical settings~\cite{RN68,abou2020preclinical}. 
Thus, we evaluated the impact of spatial intra-lesion heterogeneity on the performance of the LC-QSPECT method.

\begin{figure}
    \centering
    \includegraphics[width=0.45\textwidth]{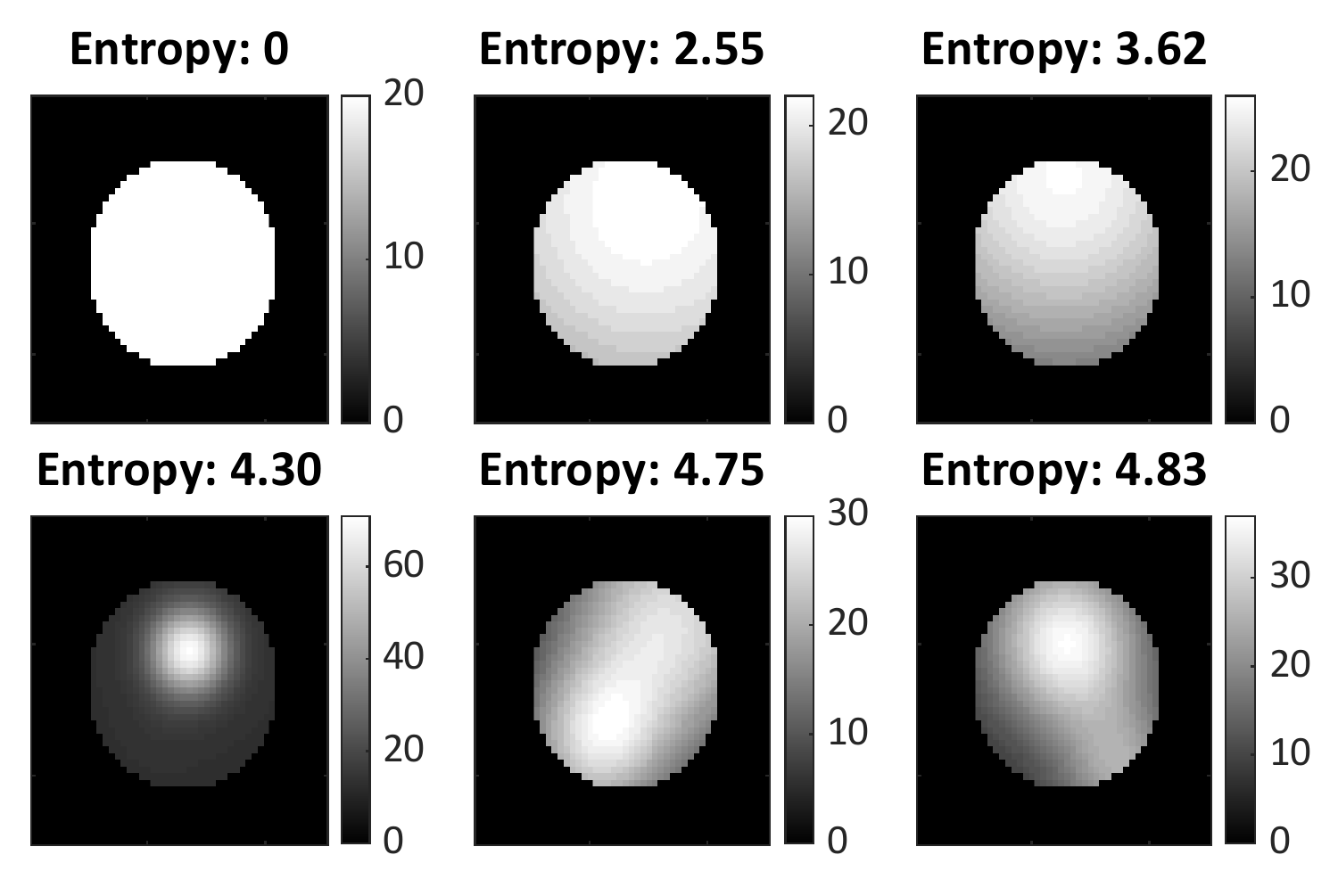}
    
    \caption{Lesions with different degrees of spatial heterogeneity.}
    \label{fig:het_lesion_img}
\end{figure}

We generated five phantoms with average patient size and a lesion of diameters 33.75~mm~\cite{RN65} in the pelvis. Each phantom had a different amount of spatial intra-lesion heterogeneity.
To simulate intra-lesion heterogeneity, we modeled the uptake in the lesion as a 3D lumpy model~\cite{liu2021observer}. Denote the support of the lesion in the object space by $s(\bm{r})$. Then, the lesion activity uptake, denoted by $f_l(\bm{r})$, is given by
\begin{equation}
    f_l(\bm{r}) = s(\bm{r})\sum^{P}_{p=1}\frac{a_p}{2 \pi \sigma^2_p} \exp \left( -\frac{|\bm{r}-\bm{c}_p|^2}{\sigma_p^2} \right),
\end{equation}
where $P$ denotes the total number of lumps, and $\bm{c}_p$, $a_p$, and $\sigma_p$ denote the center, magnitude, and width of the $p^{th}$ lump function, respectively.
Different levels of heterogeneity were simulated by varying the values of $P$, $\bm{c}_p$, $a_p$, and $\sigma_p$.
The heterogeneity was characterized by entropy, where a higher value of entropy refers to more heterogeneity in uptake. We used Shannon entropy~\cite{wu2013local}, a commonly used method to quantify entropy. As described in~\cite{pharwaha2009shannon}, we first calculated the histogram of the activity map of the lesion region. The histogram had 256 bins $\{b_w,w=0,1,...,255\}$. Denote $v$ as the total number of voxels in the lesion region, the normalized histogram $\{B_w,w=0,1,...,255\}$ was calculated using the expression $B_w = b_w / v$. Then, the entropy was computed as follows~\cite{pharwaha2009shannon}:
\begin{equation}
    E= - \sum_{w=0}^{255} B_w \mathrm{log}_2 (B_w).
\end{equation}
The generated lesions and the corresponding entropy are shown in Fig.~\ref{fig:het_lesion_img}. All lesions had the same mean activity uptake and the mean activity uptake in the background, bone, gut, and lesion region was in the ratio of 2:5:25:20. We generated 50 noise realizations for each phantom.
Using these data, we evaluated the accuracy and precision of the proposed method as a function of different levels of spatial intra-lesion heterogeneity.

\subsubsection{Evaluating effect of compensating for stray-radiation-related noise}
A key feature of the proposed method is compensating for stray-radiation-related noise. To evaluate the impact of compensating for this noise on quantification performance in $\alpha$-RPT, a comparative test was performed using the VCT setup. The LC-QSPECT method was modified to not compensate for stray-radiation-related noise by setting the term $\psi = 0$ in Eq.~\eqref{eq:MLEM}. This was compared to the proposed LC-QSPECT method that compensated for this noise.

\subsection{Evaluation using physical-phantom studies}
\label{sec:phy_pht}
\begin{figure}
    \centering
    \subfigure[]{
    \includegraphics[width=0.23000\textwidth]{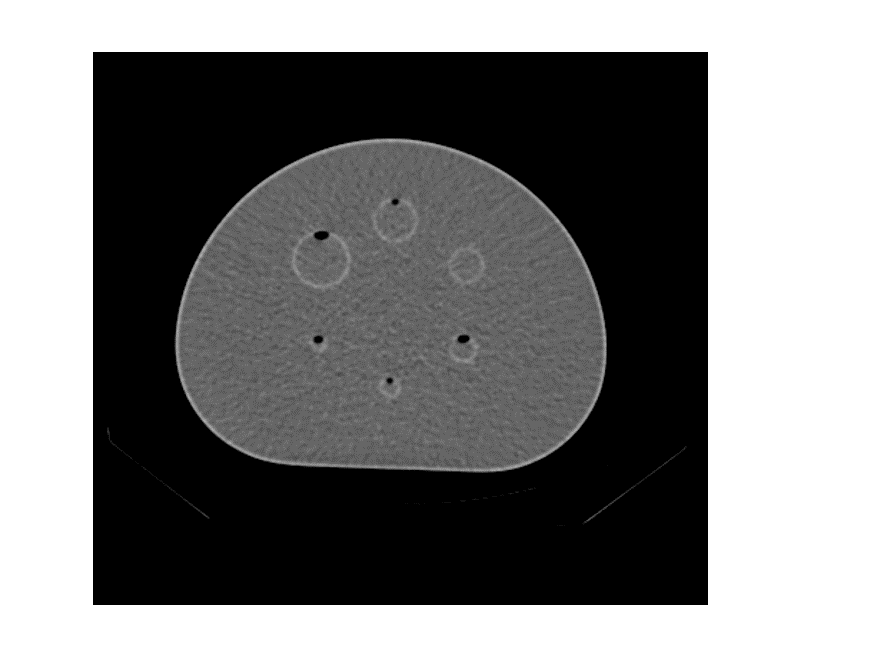}}
    \subfigure[]{
    \includegraphics[width=0.23000\textwidth]{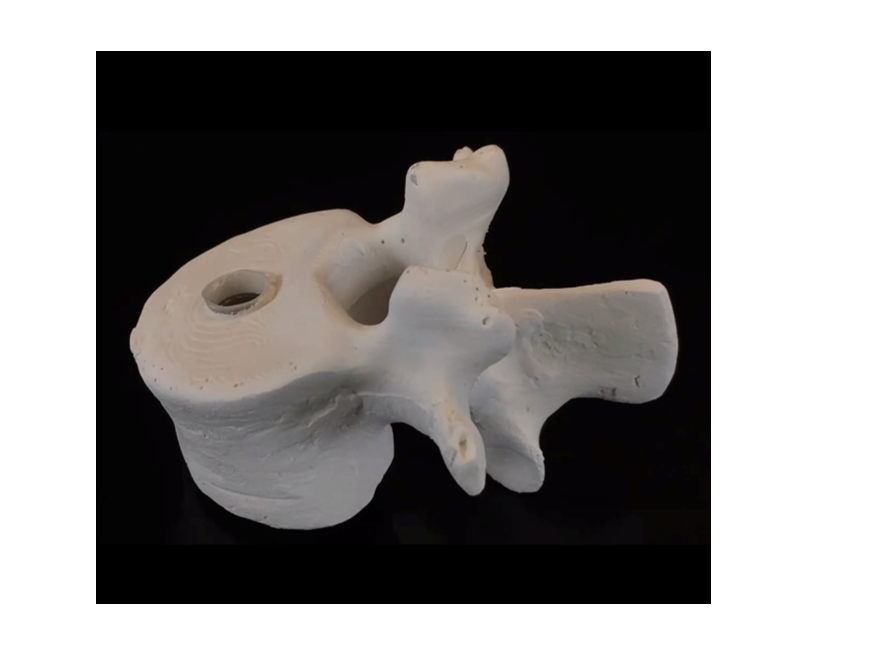}}
    \caption{(a) A CT image of the NEMA phantom and (b) a photo of the vertebrae phantom.}
    \label{fig:vert_pht}
\end{figure}

Evaluation with physical-phantom studies quantifies the performance of the method with real scanner data. 
We conducted this study with two phantoms: A NEMA phantom (Data Spectrum$^\mathrm{TM}$, USA) (Fig.~\ref{fig:vert_pht}a) and a 3D printed vertebrae phantom (Fig.~\ref{fig:vert_pht}b). 
The NEMA-phantom study was conducted to evaluate the performance of the LC-QSPECT method for different lesion sizes, with the spheres in the phantom simulating lesions. 
The vertebrae phantom study was conducted to simulate the imaging of a lesion within the spine bone in the thoracic region. 
Details on the designing, printing, and preparation of this phantom for the experiments are provided in the Supplementary material. 

\subsubsection{Phantom scanning}
\label{sec:phantom_scan}
We filled both the phantoms with clinically-relevant $\mathrm{^{223}Ra}$ activity concentrations as described in the Supplementary material. 
Next, we scanned the NEMA phantoms on a GE Optima 640 SPECT/CT system with a HEGP collimator and the vertebrae phantom in the same system with a medium energy general purpose (MEGP) collimator, with the goal of evaluating the robustness of our method for different collimator configurations. We placed each phantom at the center of the field of view of the $\gamma$-camera. 
The photopeak and scatter windows were set as 85~keV~$\pm$~20\% and 57~keV~$\pm$~18\%, respectively.
Scans were acquired at 60 angular positions spaced uniformly over 360$^{\circ}$. The acquisition time at each angular position was set to 30~seconds, as in clinical studies. The size of the projection at each angular position was 128 $\times$ 128 pixels, where the pixel side-length was 4.4~mm. A body-contour orbit was used to improve resolution. Corresponding low-dose CT scans were also acquired for each phantom (120~kVp, 10~mA, 512 x 512). The axial pixel spacing of the CT images was 0.98~mm and the spacing between slices was 5.0~mm. The CT and SPECT scans were registered. 

\subsubsection{Regional uptake estimation}
The NEMA phantom had seven VOIs, including the six spheres and the background. The vertebrae phantom had two VOIs: the lesion chamber and the background. 
VOI definitions were obtained by manually segmenting fused CT scans and OSEM reconstructed SPECT images of both phantoms. 
The segmented VOI masks had a 5~mm distance between adjacent slices, which was the same as the CT scans. We performed a cubic interpolation along the depth dimension to generate VOI masks with a 1~mm distance between adjacent slices, which gave a more accurate definition of VOI boundaries along the depth dimension.
Using these VOIs, the system matrices for the LC-QSPECT method were generated using the process as described in Sec.~\ref{sec:implementation}. From the projection data, the regional uptake was estimated using the LC-QSPECT method. In addition, we compared the performance of the LC-QSPECT method on the task of estimating the regional uptake of the NEMA phantom with and without compensating for the stray-radiation-related noise. 

The activity image was next reconstructed on the clinical workstation XELERIS (General Electric, USA) using the scanner-based OSEM technique with the following parameters: 2 iterations, 10 subsets, and a Butterworth filter with cut-off frequency of 0.48~cycle/cm and order of 10. Attenuation and scatter were compensated using the CT-based attenuation map and the dual-energy window (DEW) scatter-compensation method, respectively~\cite{jaszczak1984improved}. From the reconstructed image, the regional uptake was estimated, yielding the output with the OSEM-based method. Finally, the GTM-based method was applied to the reconstructed image, using the process described in Sec.~\ref{sec:methods_compared}. 
We computed the normalized difference between the estimated and true regional uptake, termed as the normalized error, for all three methods.


\section{Results}
\subsection{Realism of the SPECT simulation}
\label{sec:res_realism}
\begin{figure*}
    \centering
    \includegraphics[width=0.90\textwidth]{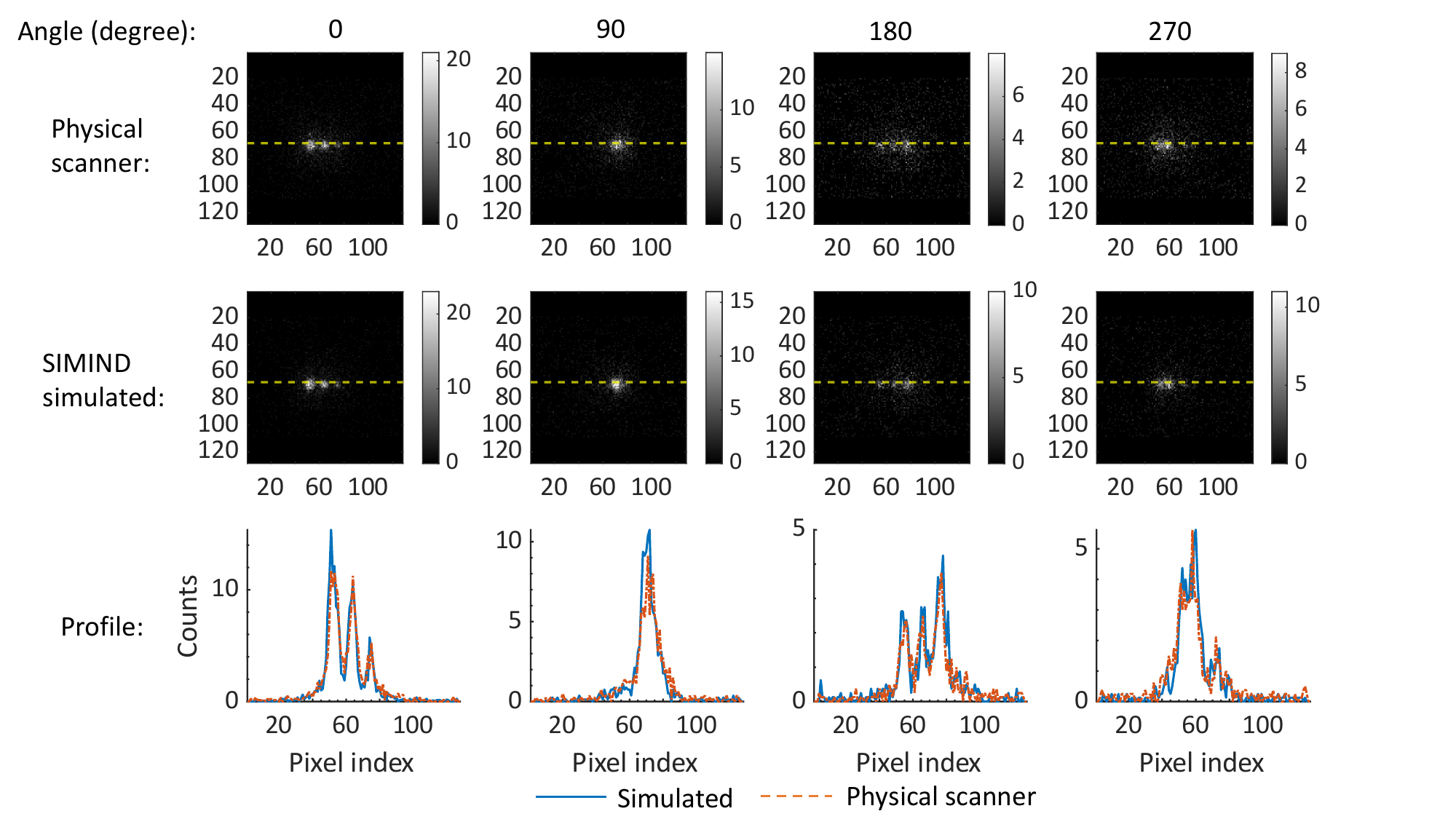}
    \caption{Projections of the NEMA phantom and the profiles along the yellow dashed lines in the projections acquired at the four angular positions using the scanner and simulated SPECT system.}
    \label{fig:profile}
\end{figure*}

Projections of the NEMA phantom from both the simulated and physical SPECT systems are shown in Fig.~\ref{fig:profile}. We compared the projections at four angular positions spaced uniformly over 360$^{\circ}$.
For each angular position, the profiles along the dashed line in both projections are also shown. For each point in the profile, the number of counts was calculated by averaging among in total 8 adjacent pixels on both sides of the dashed line to reduce the noise-related variation of the profile. We observe that the profiles of the simulated projection match those acquired on the scanner at all angular positions. This provided evidence for the realism of the SPECT system simulation in this study.


\subsection{Convergence of the proposed technique}
\label{sec:converge_result}
\begin{figure}
    \centering
    \includegraphics[width=0.40000\textwidth]{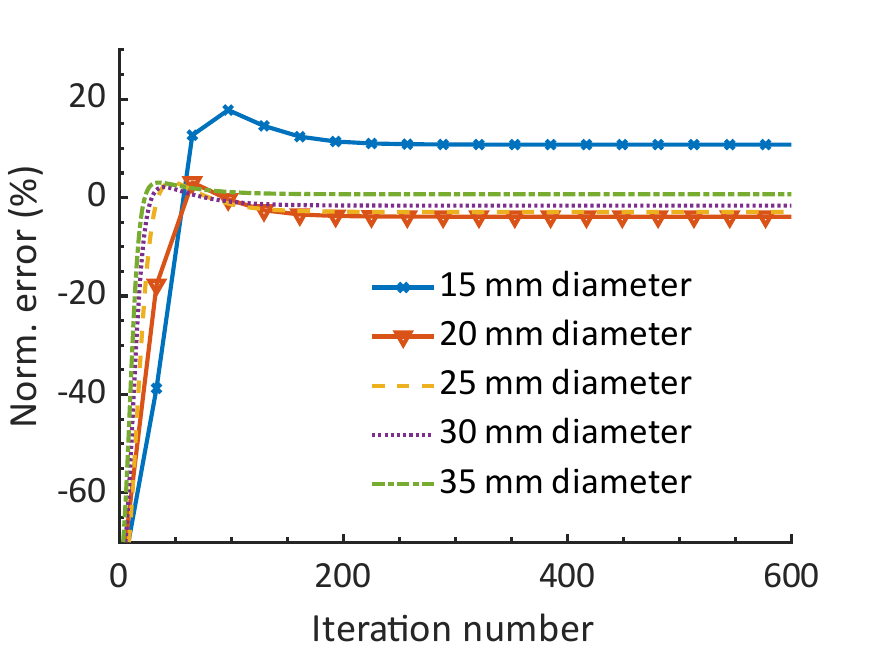}
    \caption{Normalized error in the estimated activity uptake in the lesion region with five different diameters as a function of the iteration number of the LC-QSPECT method.}
    \label{fig:converge}
\end{figure}

The normalized error in the estimated lesion uptake as a function of the iteration number of the LC-QSPECT method is shown in Fig.~\ref{fig:converge}. We observed that the method converged after 256 iterations for all five lesion diameters. Thus, we chose 256 as the number of iterations for the LC-QSPECT method for subsequent experiments. 

\subsection{Realistic simulation studies}

\subsubsection{Virtual clinical trial}
\begin{figure}
    \centering
    \subfigure[]{
    \includegraphics[width=0.23000\textwidth]{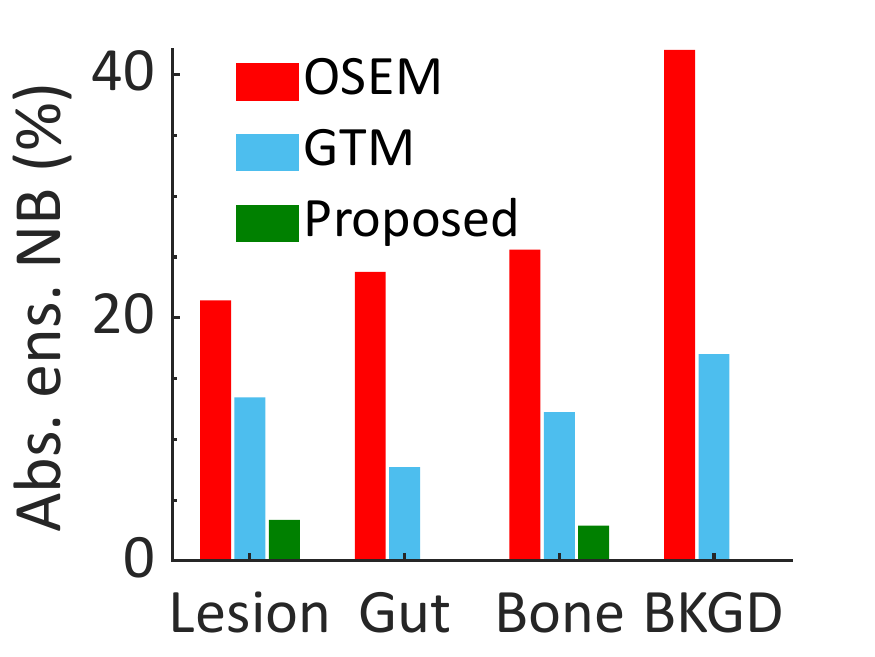}}
    \subfigure[]{
    \includegraphics[width=0.23000\textwidth]{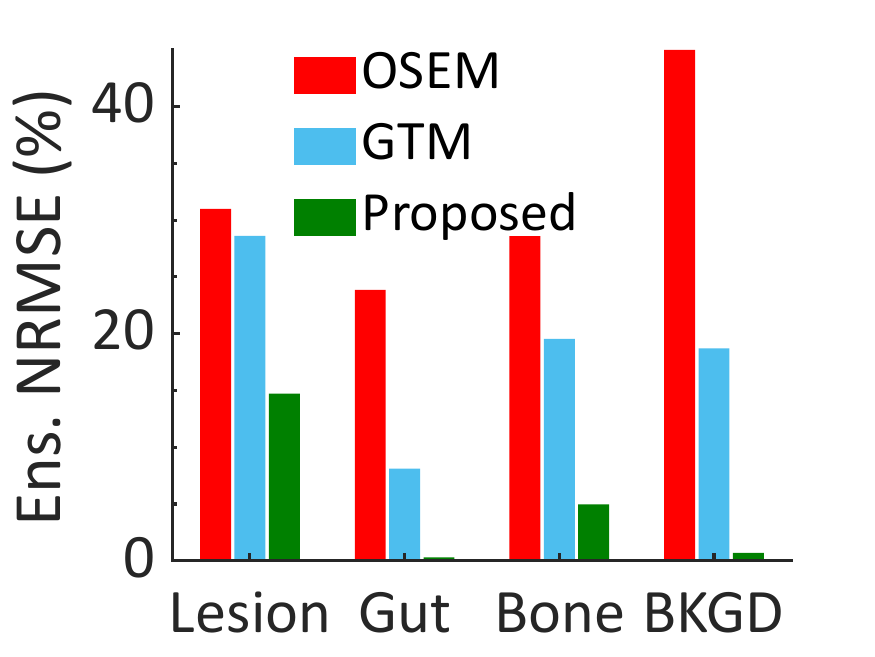}}
    \subfigure[]{
    \includegraphics[width=0.41000\textwidth]{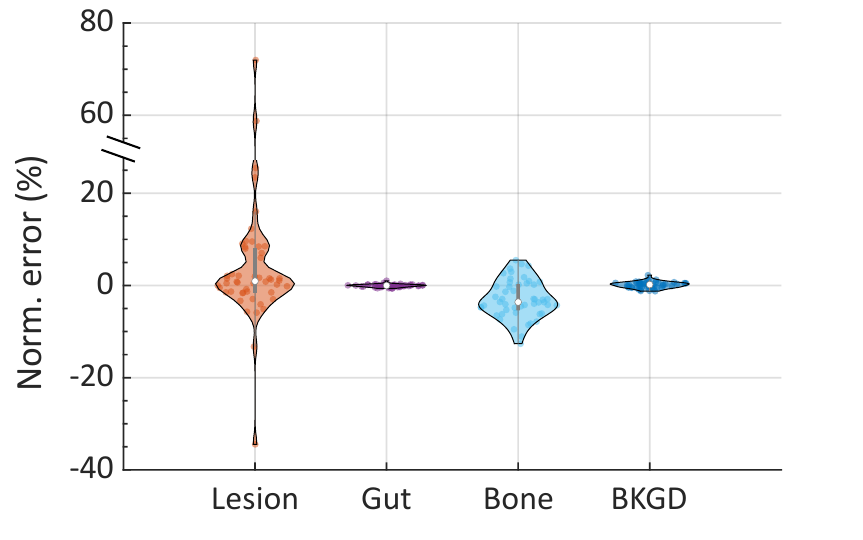}}
    
    \caption{The (a) absolute ensemble NB and (b) ensemble NRMSE of the estimated uptake across different regions using the different methods in the VCT simulating a $\mathrm{^{223}Ra}$ imaging study. (c) A violin plot showing the distribution of the normalized error (normalized by dividing by the true value) using the proposed method across all fifty patients.}
    \label{fig:estRegion}
\end{figure}

The absolute ensemble NB and ensemble NRMSE of the estimated regional uptake in the lesion, gut, bone, and background (BKGD) in the VCT setup are shown in Fig.~\ref{fig:estRegion}. Also provided is a violin plot that shows the distribution of the normalized error between the true and estimated uptake for all 50 patients using the proposed method. We observe that for all the regions, the LC-QSPECT method consistently outperformed the OSEM and GTM-based methods, based on both accuracy and overall error.
The LC-QSPECT method had at least three times lower ensemble bias in the estimated lesion uptake compared to the OSEM and GTM-based methods. Further, the absolute ensemble NB obtained with the LC-QSPECT method for all the regions was consistently lower than 3.4\%. Finally, for 90\% of the simulated patients, the normalized error in estimating the lesion uptake with the proposed method was within $\pm$ 20\%.

\subsubsection{Comparing precision of estimated regional uptake using LC-QSPECT method with Cram\'er-Rao Bound}
\begin{figure}
    \centering
    \includegraphics[width = 0.40000\textwidth]{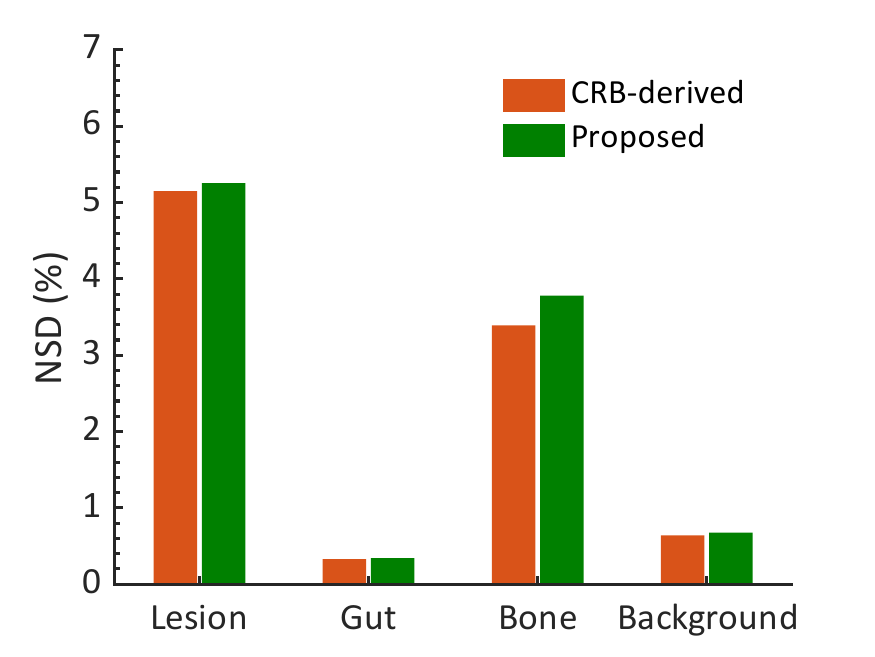}
    \caption{Comparing the NSD of the regional activity estimates obtained from the LC-QSPECT method with the lower bound for the normalized standard deviation as defined by the CRB.}
    \label{fig:CRB}
\end{figure}

Table~\ref{tb:NB_CRB} shows the NB of the estimated regional uptake from 50 noise realizations for a representative patient using the LC-QSPECT method. The NB values are all close to 0\%, with the maximum value being below 4\%. Based on these and other results as presented later, we observe that the LC-QSPECT method is approximately unbiased. Thus in
Fig.~\ref{fig:CRB}, the NSD of the corresponding estimated regional uptake using the LC-QSPECT method was compared with that obtained by taking the square root of the normalized CRB, the lower bound of variance for any unbiased estimator. We observe that for all the regions, the proposed method yielded a NSD very close to that obtained from the CRB.
\begin{table}
\centering
    \caption{NB of the estimated regional uptake for one patient using the LC-QSPECT method.}
    \begin{adjustbox}{center}
    \begin{tabular}{ccccc}
    \hline
    \T \B VOI     & Lesion & Gut   & Bone & Background \\ \hline
    \T \B NB (\%) & 0.30   & -3.08 & 0.04 & 0.28       \\ \hline
    \end{tabular}
    \end{adjustbox}
    \label{tb:NB_CRB}
\end{table}

\subsubsection{Performance as a function of lesion size and contrast}
\begin{figure}
    \centering
    \subfigure[]{
    \includegraphics[width=0.23000\textwidth]{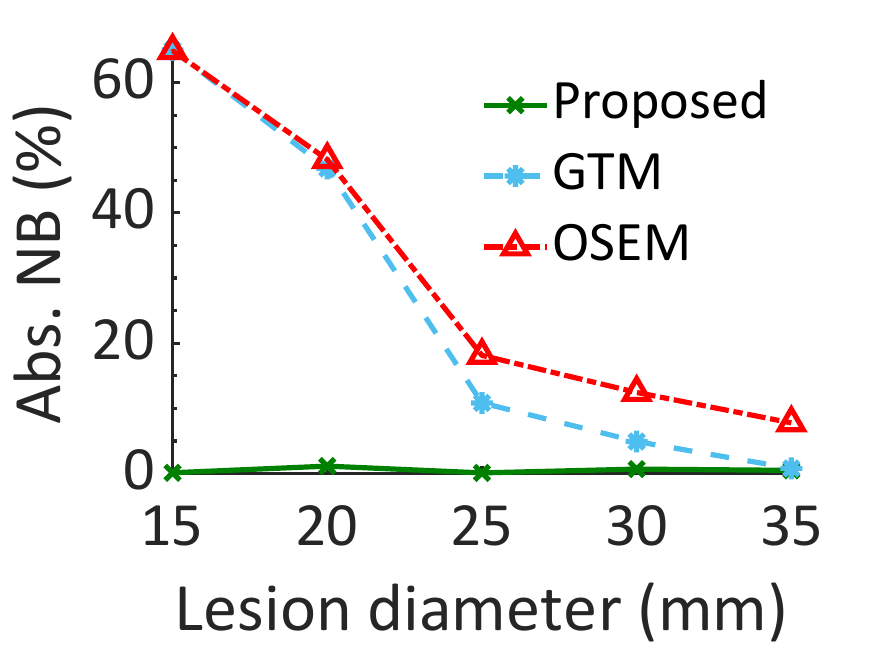}}
    \subfigure[]{
    \includegraphics[width=0.23000\textwidth]{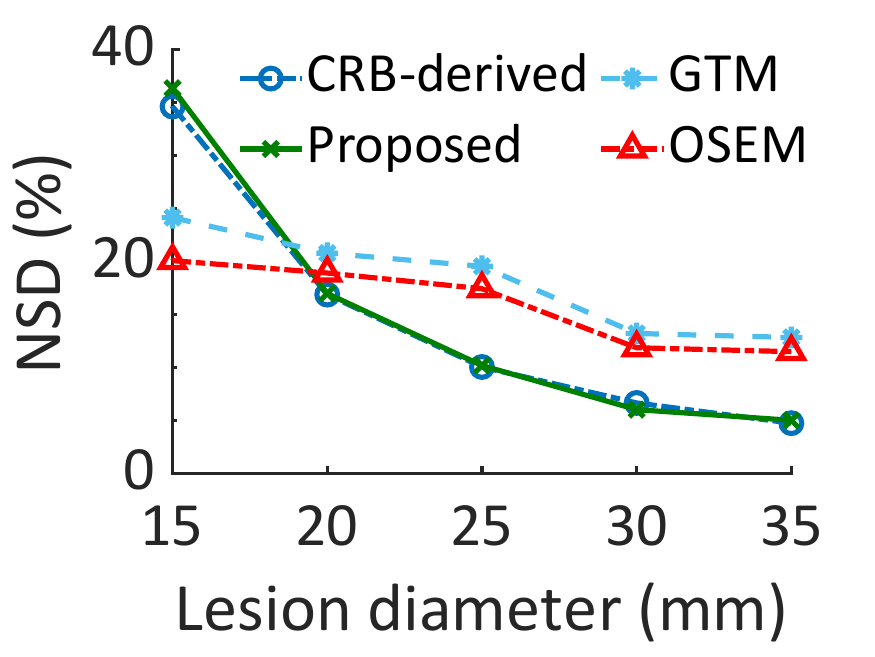}}
    \subfigure[]{
    \includegraphics[width=0.23000\textwidth]{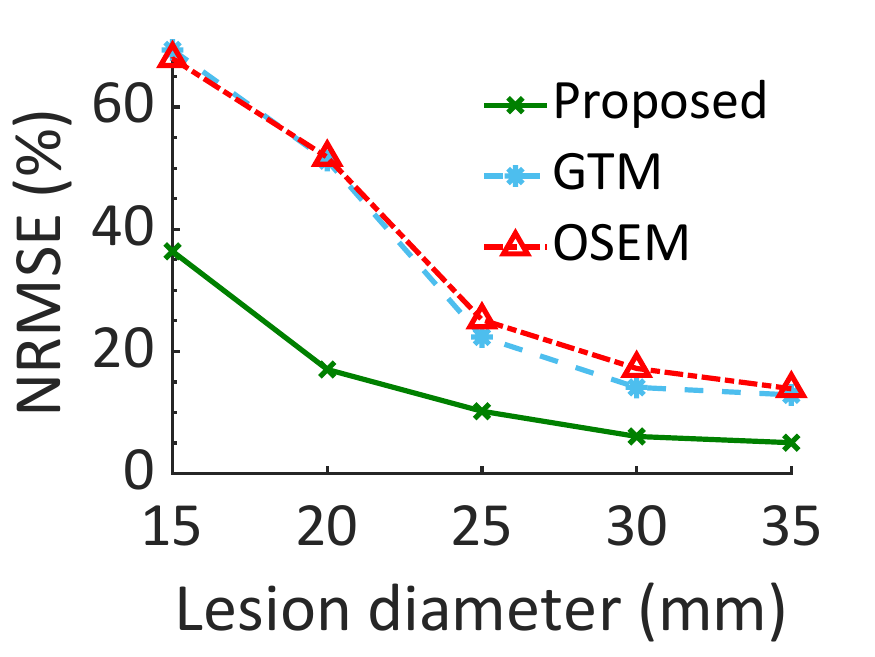}}
    \caption{The (a) absolute NB (b) NSD and (c) NRMSE between the true and estimated lesion uptake as a function of the lesion diameter in the realistic simulation study.}
    \label{fig:resSize}
\end{figure}

The absolute NB, NSD, and NRMSE of the estimated lesion uptake as a function of the lesion diameter using the different methods are shown in Fig.~\ref{fig:resSize}. The NSD of the estimates is also compared with the square root of the CRB.
The LC-QSPECT method consistently yielded close to zero bias. Further, the NSD of the estimates obtained with this method were close to that derived from the CRB for all lesion diameters. 
Additionally, the method yielded the lowest NRMSE and consistently outperformed both the OSEM and GTM-based methods for all lesion sizes.

\begin{figure}
    \centering
    \subfigure[]{
    \includegraphics[width=0.23000\textwidth]{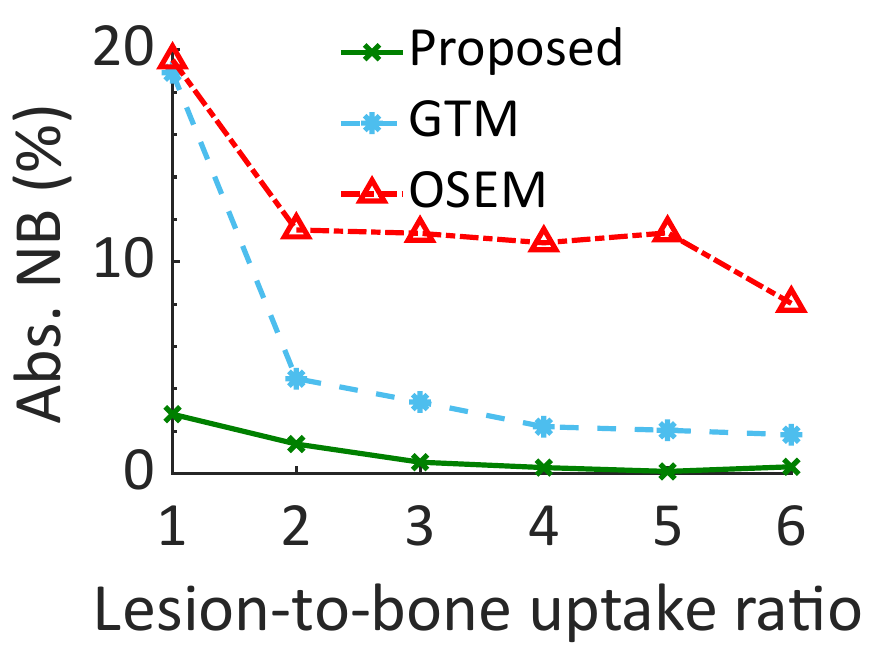}}
    \subfigure[]{
    \includegraphics[width=0.23000\textwidth]{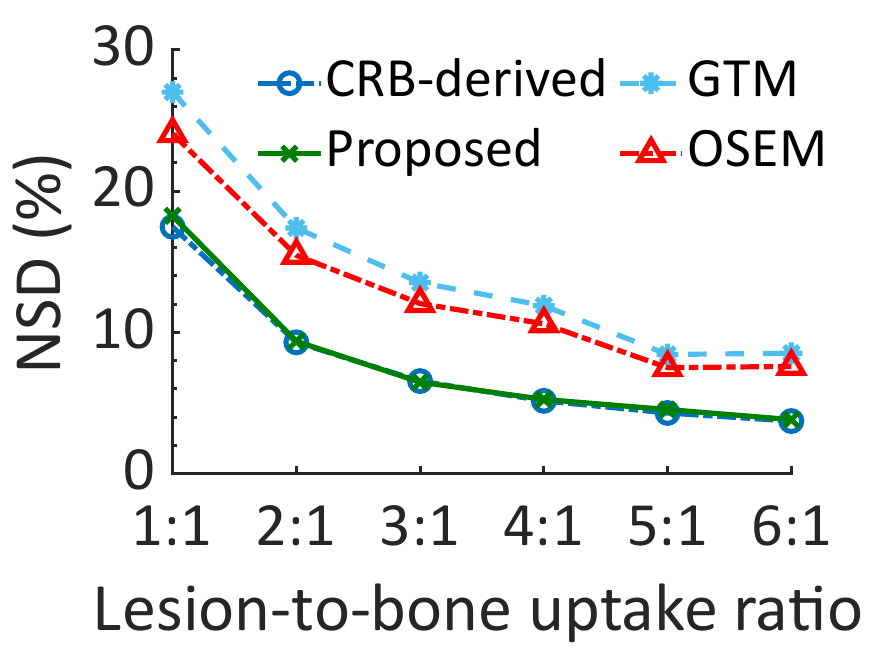}}
    \caption{The (a) absolute NB and (b) NSD of the estimated lesion uptake as a function of the lesion-to-bone uptake ratio in the realistic simulation study.}
    \label{fig:resRatio}
\end{figure}

The absolute NB and NSD of the estimated lesion uptake as a function of the lesion-to-bone uptake ratio (LBUR) using the LC-QSPECT, OSEM, and GTM-based methods are shown in Fig.~\ref{fig:resRatio}. We also compared the NSD of the estimates with that derived from the CRB.
Again, the LC-QSPECT method consistently yielded almost zero bias and NSD close to the square root of CRB for all LBUR values. Further, the method consistently outperformed both the RBQ methods.

\subsubsection{Impact of spatial intra-lesion heterogeneity}
\begin{figure}
    \centering
    \subfigure[]{
    \includegraphics[width=0.23000\textwidth]{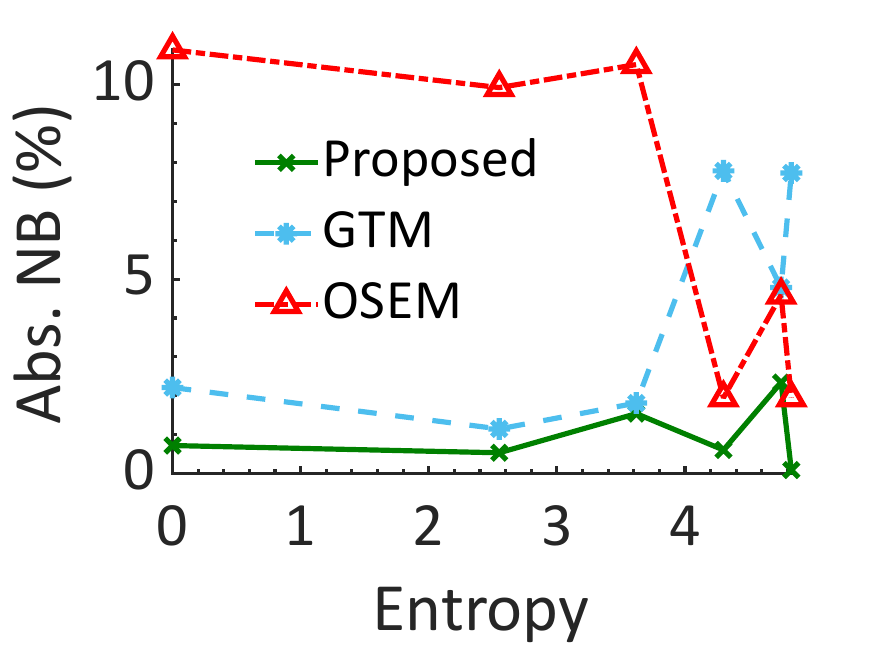}}
    \subfigure[]{
    \includegraphics[width=0.23000\textwidth]{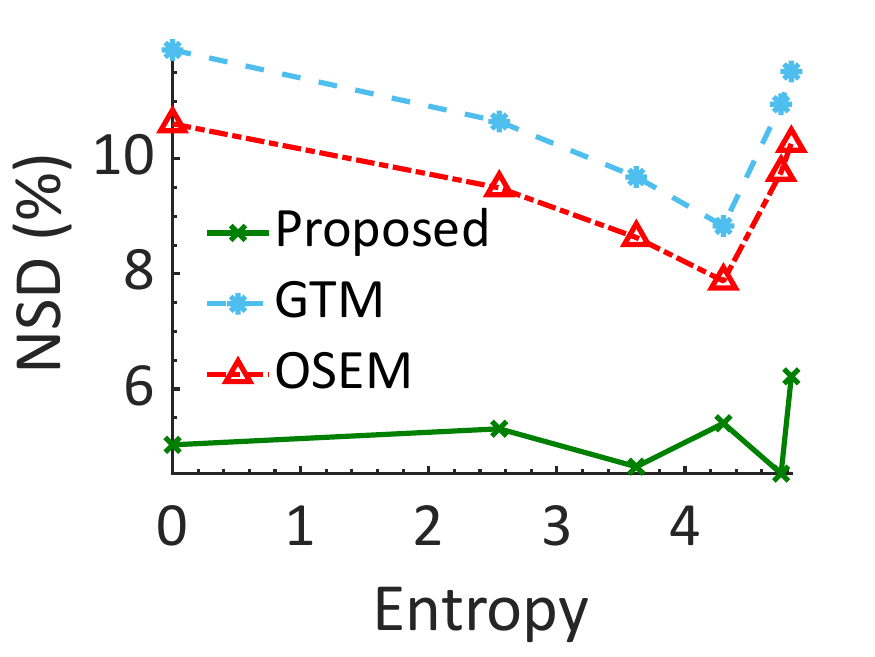}}
    \caption{The (a) absolute NB and (b) NSD of the estimated lesion uptake as a function of different amounts of intra-lesion heterogeneity as quantified using the entropy parameter.}
    \label{fig:resHete}
\end{figure}

The absolute NB and NSD of the estimated lesion uptake for different degrees of spatial intra-lesion heterogeneity, as quantified with the entropy parameter, using the LC-QSPECT, OSEM, and GTM-based methods are shown in Fig.~\ref{fig:resHete}. We observe that the LC-QSPECT method consistently yielded close to zero bias for all degrees of heterogeneity. Further, even the precision of the estimated mean uptake was not significantly impacted by the intra-lesion heterogeneity. Finally, based on both the accuracy and precision, the proposed method outperformed both the RBQ methods.

\subsubsection{Impact of compensating for stray-radiation-related noise}
The ensemble NRMSE in estimating regional uptake using the LC-QSPECT method with and without compensating for the stray-radiation-related noise in the VCT is shown in Table~\ref{tab:BGN}. We observe that compensating for the stray-radiation-related noise led to significantly more reliable regional uptake estimates in a clinically realistic $\alpha$-RPT data acquisition setup. 

\begin{table*}
\centering
\caption{Ensemble NRMSE of estimated regional uptake with the data acquired in the VCT using the proposed LC-QSPECT method and the LC-QSPECT method that did not compensate for stray-radiation-related noise.}
\begin{adjustbox}{center}
\begin{tabular}{ccccc}
\hline
\multicolumn{5}{c}{\T \B Ensemble normalized root mean square error (NRMSE) (\%)}                                                                                              \\ \hline
\T \B VOI                                                                                           & Lesion  & Gut   & Bone   & Background \\ \hline
\begin{tabular}[c]{@{}c@{}} \T \B Proposed LC-QSPECT method \\ (compensated for stray-radiation-related noise)\end{tabular} & 14.72  & 0.30  & 4.98  & 0.69      \\
\begin{tabular}[c]{@{}c@{}} \T \B LC-QSPECT method without\\compensating for stray-radiation-related noise\end{tabular}       & 187.49 & 1.11 & 29.16 & 28.26     \\ \hline
\end{tabular}
\end{adjustbox}
\label{tab:BGN}
\end{table*}

\subsection{Results of physical-phantom studies}
\subsubsection{NEMA phantom study}
\begin{figure}
    \centering
    \includegraphics[width=0.45\textwidth]{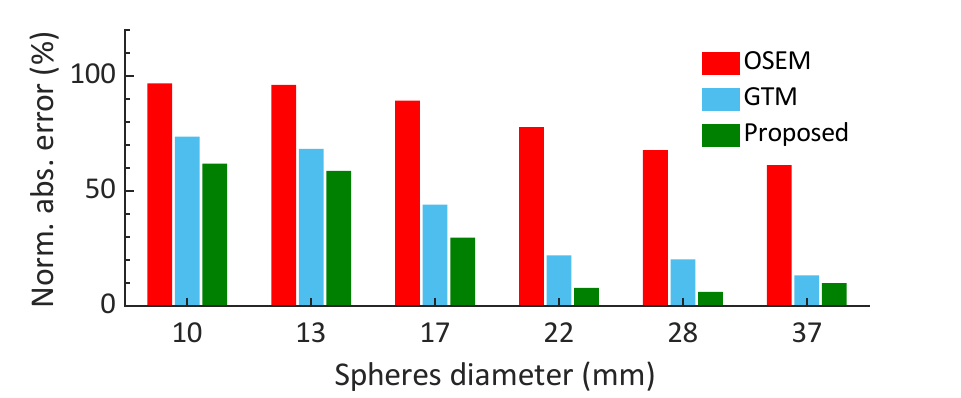}
    \caption{Normalized absolute error of regional uptake estimates of the NEMA phantom using the LC-QSPECT, GTM, and OSEM-based methods.}
    \label{fig:nema_bias}
\end{figure}

The normalized absolute error in estimating the regional uptake as a function of the sphere diameter is shown in Fig.~\ref{fig:nema_bias}. We observe that the LC-QSPECT method consistently outperformed both the RBQ methods. Next, the normalized error of the LC-QSPECT method with and without compensating for the stray-radiation-related noise is shown in Table~\ref{tb:NEMA_srn}. We observe that the values of absolute normalized error in the largest three spheres of the NEMA phantom are at least three times higher than that of the LC-QSPECT that compensates for the stray-radiation-related noise.
\begin{table*}
\centering
    \caption{Normalized error of estimated regional uptake in the NEMA phantom using the LC-QSPECT method with and without compensating for the stray-radiation-related noise.}
    \begin{adjustbox}{,center}
    \begin{tabular}{ccccccc}
    \hline
    \multicolumn{7}{c}{ \T \B Normalized error (\%)}                                                     \\ \hline
    \begin{tabular}[c]{@{}c@{}} \T \B Sphere diameter (mm)\end{tabular} & 10 & 13 & 17 & 22 & 28 & 37 \\ \hline
    \begin{tabular}[c]{@{}c@{}} \T \B LC-QSPECT method \\ (compensated for stray-radiation-related noise)\end{tabular}        & -62.02 & -58.86  & -29.84 & -7.95 & 6.28 & -10.03 \\
    \begin{tabular}[c]{@{}c@{}} \T \B LC-QSPECT method without \\ compensating for stray-radiation-related noise\end{tabular} & 61.43  & 47.97 & 55.31 & 37.63  & 41.20  & 34.31  \\ \hline
    \end{tabular}
    \end{adjustbox}
    \label{tb:NEMA_srn}
\end{table*}

\subsubsection{Vertebrae phantom study} 
The normalized absolute error in estimating the lesion uptake using the LC-QSPECT, OSEM, and GTM-based methods is shown in Table~\ref{tb:vert_bias}. Again, the LC-QSPECT method significantly outperformed the other two approaches. 
Further, the LC-QSPECT method yielded an error as low as 2.86\% in these challenging real-world conditions.
\begin{table}
\centering
    \caption{Normalized error of the estimated lesion uptake for the vertebrae phantom.}
    \begin{adjustbox}{center}
    \begin{tabular}{c c c c}
    \hline
    \T \B Methods& LC-QSPECT &GTM &OSEM\\[0.5ex] 
    \hline
    \T \B Normalized error~(\%) & -2.86 & -13.80 & -65.85\\[0.5ex] 
    \hline
    \end{tabular}
    \end{adjustbox}
    \label{tb:vert_bias}
\end{table}
\section{Discussion}
We have designed, implemented, and evaluated an LC-QSPECT approach to quantify regional activities from low-count SPECT data for $\alpha$-RPTs.
Our results demonstrate the efficacy of the LC-QSPECT method for $\alpha$-RPTs. In Fig.~\ref{fig:estRegion}, we observed that the LC-QSPECT method yielded averaged absolute ensemble NB and ensemble NRMSE values as low as 1.6\% and 5.2\%, respectively. In contrast, the OSEM-based method yielded averaged absolute ensemble NB and ensemble NRMSE of 28.2\% and 32.1\% respectively, and the GTM-based method yielded averaged absolute ensemble NB and ensemble NRMSE of 12.6\% and 18.7\%, respectively. These observations for the OSEM-based method are consistent with previous literature~\cite{RN21,RN60,RN61} and confirm the limitation of this approach for $\alpha$-RPTs. Further, our results demonstrate the reliability and superiority of the proposed method for $\alpha$-RPTs. 

Figs.~\ref{fig:CRB},~\ref{fig:resSize}b,~and~\ref{fig:resRatio}b show that the NSD of the regional uptake estimates obtained using the LC-QSPECT method approached the lowest theoretical limit as defined by the CRB. 
This is an important finding as quantifying uptake precisely is a major challenge in $\alpha$-RPT SPECT~\cite{RN61}.
Further, our results in Figs.~\ref{fig:estRegion}a,~\ref{fig:resSize}a,~\ref{fig:resRatio}a, and Table~\ref{tb:NB_CRB} indicate that the LC-QSPECT method is approximately unbiased. These findings, while empirical, are theoretically consistent as the LC-QSPECT method is an ML estimator, and when an efficient estimator exists, the ML estimator is efficient~\cite{RN50}. Thus, our results suggest that the LC-QSPECT method may be the optimal estimator in terms of bias and variance properties. 
In Fig.~\ref{fig:resSize}b we observe that for certain lesion sizes the RBQ methods yielded a lower NSD than the proposed method and that derived from CRB. However, we also observe that the RBQ methods are biased for all the lesion sizes considered in this experiment, so the comparison of the variance of these methods with CRB is not meaningful. Further, we see that even for those lesion sizes, the proposed method yielded an improved overall accuracy and precision as quantified by the NRMSE.


The efficacy of the LC-QSPECT method was also observed for different lesion sizes (Fig.~\ref{fig:resSize}) and different LBURs (Fig.~\ref{fig:resRatio}). The bias in the estimated uptake was close to zero for all the considered lesion sizes and LBUR values. In particular, the low bias at different lesion sizes demonstrates that the LC-QSPECT method is relatively insensitive to PVEs. PVEs are a major source of error in quantitative SPECT. Note that even though the GTM-based method is designed exclusively to compensate for PVEs, the method still yielded large values of bias for small lesions. In contrast, our method was approximately unbiased~(Fig.~\ref{fig:resSize}).
This result thus demonstrates an important advantage of this method compared to OSEM and GTM-based methods.
Similarly, the proposed method was approximately unbiased for different LBUR values, demonstrating the accuracy of the method for different signal contrast. 

While the LC-QSPECT method assumes that the activity uptake in each VOI is homogeneous, the results in Fig.~\ref{fig:resHete} show that the intra-lesion heterogeneity evaluated in this study does not substantially affect the reliability of the estimated uptake.
Further, very importantly, even when such heterogeneity is present, the proposed method outperforms the OSEM and GTM-based approaches.
Thus, even when the assumption of uniform uptake in the lesion is violated to some extent, these results suggest that the LC-QSPECT method can still be a reliable and superior option.

In the physical-phantom studies, we again observed that the LC-QSPECT method significantly outperformed both the RBQ methods (Fig.~\ref{fig:nema_bias} and Table~\ref{tb:vert_bias}). 
These results were consistent with the simulation-study results.
More importantly, they also demonstrate the feasibility of the proposed method in real-world settings. The LC-QSPECT method requires the definitions of VOIs, and the physical-phantom studies show that these can be obtained from segmenting fused CT and reconstructed SPECT images. CT scans are typically acquired in conjunction with the SPECT scans for attenuation compensation, and often using SPECT/CT systems. 

We chose a different collimator configuration for the vertebrate and the NEMA phantom in the physical-phantom studies, with the goal of evaluating the method for different collimator configurations and thus assessing the robustness of the method. We observed that the LC-QSPECT method yielded more accurate estimates for both phantoms than the RBQ methods. Given this result, a future area of research is evaluating the performance of the method for a larger number of scanner and collimator configurations.

An additional challenge for quantitative $\alpha$-RPT SPECT is the large proportion of stray-radiation-related noise. As shown in Table~\ref{tab:BGN}, the impact of compensating for this noise while performing the quantification was significant, especially for the lesion, where, not compensating for this noise resulted in over 180\% NRMSE.
We observed that the proposed LC-QPSECT method effectively compensated for this stray-radiation-related noise. 
Similar results were observed in the physical-phantom study where compensating for the stray-radiation-related noise improved the estimation accuracy of the proposed method for the four larger spheres in NEMA phantom, as shown in Table~\ref{tb:NEMA_srn}.
For the smallest two spheres, however, the absolute values of normalized error without stray-radiation-related noise compensation using the LC-QSPECT method are slightly lower than those with this compensation. Here we note that for the LC-QSPECT method, VOIs are defined from manual segmentation of the low-dose CT scans. These VOIs may have errors. For the small sphere sizes, even a small mis-definition of the VOI may cause a large error. Further, these results are only for a single noise realization and the impact of stochastic noise-related errors is high for smaller spheres. Thus, it is likely that, for these smaller spheres, the overestimation caused due to the stray-radiation-related noise was canceled by the underestimation due to the noise-related stochastic error and the potential mis-definition of the VOI maps.
However, the average values of absolute normalized error among all spheres in the NEMA phantom using the LC-QSPECT with and without compensating for the stray-radiation-related noise were 29.2\% and 46.3\%, respectively. This demonstrated that overall, the method yielded much improved performance. 

Another important feature of the LC-QSPECT method is the use of an MC simulation-based approach to generate the system matrix. This approach yields highly accurate modeling of SPECT physics. The MC approach is computationally feasible because the number of VOIs is typically quite small.
Thus, this matrix can be pre-computed and stored. In this study, the system matrix of each patient took less than 50~minutes and 30~MB for generation and storage, respectively. In clinical applications, the CT scans of the patient could be acquired first, then the system matrix can be generated simultaneously when the SPECT scan is acquired.
Further, the estimation process is rapid; 256 iterations required $<$ 30~seconds on a standard desktop computer equipped with an Intel(R) Core(TM) i7-10700K CPU with 16 cores and 32~GB RAM.
This is unlike developing such an approach for OSEM-based methods, 
for which a similar system matrix may require up to 30 TB of memory.
Thus, the proposed method provides a mechanism for highly accurate Monte Carlo-based system modeling, which is not possible with RBQ methods. This is another advantage of the proposed method.

A limitation of the LC-QSPECT method is that the method quantifies regional uptake rather than voxel values. Thus, dosimetry estimates are confined to the mean absorbed dose across the region. This may then preclude the estimation of a dose-volume histogram within tumors and normal organs, which is used in external-beam radiation therapy. However, given the errors with voxel-based reconstruction methods, the estimation of dose-volume histograms may be infeasible at these low counts. In contrast, our method shows that the mean regional uptake can be estimated reliably. Another point to note is that the exchange of (purported) resolution from conventional voxel-based to this region-based method is of limited consequence to $\alpha$-RPT dosimetry as no imaging system can resolve isotope distribution at the sub-100~$\mu m$ scale. 

The LC-QSPECT method, similar to the RBQ methods, requires reliable VOI definitions. While these VOI definitions could be obtained from the CT scans that are acquired for quantitative SPECT, there is a possibility that the SPECT and CT scans may be misaligned. Studying the effect of this misalignment is an important research direction. Methods to register SPECT and CT scans~\cite{scott1995image,tang2006co} would help address this limitation. 
Another area of future research is the clinical validation of these techniques. One issue with such validation is the absence of ground-truth quantitative values. To address this issue, no-gold-standard evaluation techniques are being developed~\cite{hoppin2002objective}, including in the context of evaluating quantitative SPECT reconstruction methods~\cite{jha2016no}. This may provide a mechanism to clinically evaluate the proposed methods. 

\section{Conclusion}
A low-count quantitative SPECT (LC-QSPECT) method was proposed for quantitative SPECT with $\alpha$-particle emitting radiopharmaceutical therapies ($\alpha$-RPTs), where the number of detected $\gamma$-ray photons is very small. 
The method yielded reliable (accurate and precise) values of regional uptake and outperformed the conventional OSEM and GTM-based methods, as evaluated in the context of $\alpha$-RPT with $\mathrm{^{223}Ra}$. The method yielded reliable activity estimates in a virtual clinical trial simulating imaging of patients with bone metastasis who were administered this therapy. Additionally, the method estimated reliable uptake for different lesion sizes and different lesion-to-bone uptake ratio values. Further, the method yielded reliable lesion uptake estimates for different degrees of intra-lesion heterogeneity.
The method was observed to be approximately unbiased and yield a standard deviation close to that defined by the Cram\'er-Rao bound, indicating that the method may be an efficient estimator.
Evaluation with physical-phantom studies on SPECT/CT scanners using NEMA and an anthropomorphic vertebrae phantom reinforced this reliable quantification performance and provided evidence for the feasibility of the approach in practical settings.
Overall, the results provide strong evidence for further evaluation and application of this method to perform quantitative SPECT for $\alpha$-RPTs.

\section*{Acknowledgement}
This work was supported in part by grants R21-EB024647, R01-EB031051 and R56-EB028287, awarded by National Institute of Biomedical Imaging and Bioengineering. This work was also supported by 2020 Student Research Grant, awarded by Society of Nuclear Medicine and Molecular Imaging.
We also thank the Washington University Center for High Performance Computing for providing computational resources for this project. The center is partially funded by National Institutes of Health (NIH) grants 1S10RR022984-01A1 and 1S10OD018091-01. All authors declare that they have no known conflicts of interest in terms of competing financial interests or personal relationships that could have an influence or are relevant to the work reported in this paper.

\bibliographystyle{IEEEtran} 
\bibliography{IEEEabrv,MLROI_Manu}
\end{document}


\bstctlcite{IEEEexample:BSTcontrol}
\title{Supplementary Material}
\maketitle

\section{Designing, printing, and assembling the vertebral phantom}
\subsection{Segmentation and creating 3D models}
Anonymized CT scan of an adult male (26 yrs) was used to segment lumbar vertebrae. Using segment editor module of 3D slicer, lumbar vertebrae L1-L5 were segmented. Each vertebra was exported as .stl files for further processing. Blender v2.8 was used to process the 3D models. Operations performed in Blender include cleaning the mesh, fixing normal, retopology, Laplacian smoothing, etc. After creating clean vertebral models, L4 vertebrae was chosen for creating necessary molds for phantom fabrication. A core for the L4 vertebrae was designed by scaling down the body region of the vertebrae. Thickness of 1 mm was given to the designed core. The core was designed to have a hole and a plug to enable refilling and leakproof closing. A shell was designed in two halves around the whole vertebrae with shell thickness of 1 mm.
\subsection{3D printing}
The core and two halves of the mold were 3D printed using FormLabs Form2 3D printer. The core was 3D printed using Clear resin at standard resolution. The mold and plug were printed using Elastic 50A resin at standard resolution. 3D printed parts were post-processed as suggested by the manufacture, which includes support removal, Isopropyl alcohol wash and curing.
\subsection{Assembling and fabrication of the phantom}
The core was placed inside the mold and seams of the molds were taped to hold the parts in place. The void between the core and the mold mimics the cortex of the vertebrae. This void was filled with Plaster of Paris (PoP)(DAP PoP). A slurry of PoP was prepared following the manufacturer’s instructions (2 parts powder and 1-part cold water). Using the inlet built into the mold, PoP was poured and let to set overnight. The plaster model was carefully taken out of the elastic mold and spray coated with acrylic paint.

\section{Physical phantom preparation}

We filled both the phantoms with clinically-relevant $\mathrm{^{223}Ra}$ activity concentrations. $\mathrm{^{223}Ra}$ was purified from a source of Actinium-227 and a source of Thorium-227 for the NEMA and vertebrae phantoms respectively. We diluted the $\mathrm{^{223}Ra}$ in methanol: nitric acid (2N) (80:20). For each phantom, we prepared the solution of $\mathrm{^{223}Ra}$ with an activity concentration of 40~kBq/ml by diluting with deionized water. The activity was carefully measured with the radionuclide calibrator CRC$^\circledR$ -15 Dual PET (Capintec). We filled each sphere of the NEMA phantom with this solution using a 20~ml syringe with an 18Gx6~in needle. The remaining phantom was filled with deionized water to simulate attenuation and scatter due to soft tissue. For the vertebrae phantom, we filled the lesion chamber with the $\mathrm{^{223}Ra}$ solution using a 20~ml syringe with an 18Gx3/4” needle. Then we fixed the filled vertebrae phantom on a cylindrical insert inside the NEMA phantom body. We filled the rest of the phantom with deionized water, modeling soft tissue. This configuration simulated the imaging of a lesion in the spine in the thoracic region.